\documentclass[11pt]{article}

\usepackage[final]{acl}

\usepackage{times}
\usepackage{latexsym}

\usepackage[T1]{fontenc}

\usepackage[utf8]{inputenc}

\usepackage{microtype}

\usepackage{inconsolata}

\usepackage{graphicx}
\usepackage{subcaption}
\usepackage{booktabs}
\usepackage{multirow}
\usepackage{siunitx}

\usepackage{amsmath} 

\usepackage{cleveref}

\usepackage{enumitem}

\usepackage[frozencache=true,cachedir=./]{minted2}  
\usepackage{ragged2e}

\definecolor{LightGray}{gray}{0.9}
\PassOptionsToPackage{dvipsnames}{xcolor}

\title{Rerank Before You Reason: Analyzing Reranking Tradeoffs through Effective Token Cost in Deep Search Agents}

\author{Sahel Sharifymoghaddam \and Jimmy Lin \\
  David R.\ Cheriton School of Computer Science,\\
 University of Waterloo, Canada \\
  \texttt{\{sahel.sharifymoghaddam,jimmylin\}@uwaterloo.ca} }

\begin{document}
\maketitle
\begin{abstract}
Deep research agents rely on iterative retrieval and reasoning to answer complex queries, but scaling test-time computation raises significant efficiency concerns. We study how to allocate reasoning budget in deep search pipelines, focusing on the role of listwise reranking. Using the BrowseComp-Plus benchmark, we analyze tradeoffs between model scale, reasoning effort, reranking depth, and total token cost via a novel effective token cost (ETC) metric. Our results show that reranking consistently improves retrieval and end-to-end accuracy, and that moderate reranking often yields larger gains than increasing search-time reasoning, achieving comparable accuracy at substantially lower cost. All our code is available at \url{https://github.com/sahel-sh/DeepHone}.
\end{abstract}

\section{Introduction and Related Work}
Deep research agents~\cite{huang2025deep, selfrag} combine large language models (LLMs) with iterative retrieval and reasoning to answer complex, multi-hop queries that exceed the scope of single-round retrieval-augmented generation (RAG)~\cite{lewis2020retrieval, ram2023context}. By decomposing queries into sequences of search actions and reasoning over intermediate evidence, such agents achieve strong performance on reasoning-intensive benchmarks~\cite{huang2025deep, zhang2025survey}. However, scaling test-time computation in this manner introduces substantial overhead, making deep search increasingly constrained by efficiency rather than accuracy alone~\cite{agenttts}.

Recent work has emphasized improving efficiency in LLM-based systems, particularly for long-context and multi-step reasoning. Techniques such as prompt caching~\cite{gim2024prompt} and prefix reuse reduce input costs, while methods like AgentDiet~\cite{agentdiet} reduce redundant context in agent trajectories. Nevertheless, retrieval-stage efficiency remains less understood, as many evaluations rely on opaque web search APIs that conflate retrieval quality with external service behavior~\cite{huang2025deep}.

To address this limitation, Chen et al. introduce BrowseComp-Plus~\cite{browsecompplus} for controlled analysis of deep research by fixing the retrieval stage to a human-verified document corpus. Their experimental results show that stronger retrievers can simultaneously improve effectiveness and reduce search calls, while increased reasoning effort in agents often yields gains at higher token cost. Despite this progress, the role of \emph{reranking} in deep research pipelines remains underexplored. While rerankers are standard in information retrieval systems, their interaction with agentic reasoning and their cost-effectiveness in multi-turn search have not been systematically studied.

In this work, we analyze listwise reranking~\cite{rankgpt,rankllm} in deep search using the BrowseComp-Plus setup. We introduce a simple and practical \emph{Effective Token Cost (ETC)} metric to quantify efficiency--effectiveness tradeoffs. Our results show that reranking consistently improves retrieval quality and reduces end-to-end token cost, with lightweight reasoning at this stage providing the best cost--effectiveness balance. We further extend ETC to heterogeneous settings, showing that smaller cross-encoder~\cite{monobert, monot5} rerankers retain most of the end-to-end accuracy gains of listwise reranking while further reducing ETC.

\vspace{-0.1cm}
\section{Experimental Setup}
We adopt the BrowseComp-Plus setup, where LLMs perform agentic search by iteratively generating queries and invoking a retrieval tool. We use the same search prompt as shown in \Cref{fig:search-prompt} and the same retriever, qwen3-embedding-8b~\cite{qwen3}, as in the original setup throughout.

OpenAI's gpt-oss-20b and gpt-oss-120b act as search agents under low, medium, and high reasoning settings. Unlike approaches that toggle thinking (e.g., qwen3 family) or cap thinking tokens (e.g.,  vLLM inference), these models vary reasoning effort via system prompts. Following OpenAI guidelines, maximum output lengths are set to 2k, 8k, and 16k tokens. All experiments use the full dataset of 830 deep search queries.

\paragraph{Retrieval and Reranking.}
At each iteration, the retriever returns the top-5 documents per query; these are then truncated to 512 tokens each before being passed to the search agent to control context growth.
With reranking, the top-$d$ candidates ($d \in {10, 20, 50}$) are first retrieved, truncated, then reranked, and the final top-5 are passed to the agent. For cost--effectiveness consistency, we use oss-20b and oss-120b as zero-shot listwise rerankers (prompt in~\Cref{fig:reranking-prompt}). Reranking uses RankLLM~\cite{rankllm} with window size 20; for $d=50$, a sliding window with stride 10 is applied. An ablation study uses qwen3-reranker-0.6 to evaluate a heterogeneous setup.

\paragraph{Hardware Setup.}
All inference runs on vLLM (v0.13.0) using NVIDIA H200 (141GB) or RTX PRO 6000 Blackwell Max-Q (96GB) GPUs (full list in \Cref{tab:hardware_all}). Search and reranking are executed on separate GPUs to avoid KV-cache and throughput interference. Rerankers use a 32k context length, while search agents use up to 128k with automatic truncation in multi-turn settings.

\paragraph{Evaluation Metrics.}
Retrieval is evaluated with Recall@5/10 and NDCG@5/10. End-to-end evaluation follows the BrowseComp-Plus protocol, reporting Accuracy (LLM-as-a-judge), Recall, number of Search Calls, and Calibration Error~\cite{phan2025humanity}. Accuracy is computed using oss-120b with the BrowseComp-Plus judging prompt (\Cref{fig:judging-prompt}), averaged over five evaluations per run. Due to compute limits, each configuration is run once; cross-run variance is analyzed in Appendix~\ref{app:cross-run variance}.

\section{Effective Token Cost}
To quantify the efficiency--effectiveness tradeoff, we define the \emph{Effective Token Cost} (ETC), a metric that accounts for hardware and financial cost disparities across token types:
\begin{equation}
\textrm{ETC} = \textrm{Input}_{nc} + \alpha \cdot \textrm{Input}_{c} + \beta \cdot \textrm{Output}_{t}
\end{equation}

\begin{table}[tbh]
\centering
\caption{Reranking effectiveness in a one-shot setting: the original query retrieves 100 candidates via qwen3-embedding-8b; top-$d \in \{10,20,50\}$ are reranked using oss-20b and oss-120b listwise rerankers. Relevance is judged using evidence documents.}
\vspace{-2mm}
\label{tab:retrieval_evidence}
\resizebox{\columnwidth}{!}{
\begin{tabular}{@{}lrrrrr@{}}
\toprule
\textbf{Retriever/Reranker} & \textbf{d} & \textbf{NDCG@5} & \textbf{NDCG@10} & \textbf{Recall@5} & \textbf{Recall@10} \\ \midrule
(0) qwen3-emb-8b & 0 & 19.72 & 20.74 & 14.91 & 20.77 \\ \midrule
(1a) oss-20b-low & 10 & 27.30 & 25.33 & 18.48 & 20.77 \\
(1b) oss-20b-med & 10 & 28.34 & 25.92 & 19.07 & 20.77 \\
(1c) oss-120b-low & 10 & 29.64 & 26.68 & 19.69 & 20.77 \\
(1d) oss-120b-med & 10 & 29.78 & 26.82 & 19.60 & 20.77 \\ \midrule
(2a) oss-20b-low & 20 & 32.28 & 29.85 & 22.54 & 25.09 \\
(2b) oss-20b-med & 20 & 34.37 & 31.50 & 23.85 & 26.06 \\
(2c) oss-120b-low & 20 & 35.69 & 32.30 & 24.52 & 26.28 \\
(2d) oss-120b-med & 20 & 36.63 & 33.01 & 25.13 & 26.70 \\ \midrule
(3a) oss-20b-low & 50 & 35.89 & 33.29 & 25.42 & 28.41 \\
(3b) oss-20b-med & 50 & 39.86 & 36.66 & 28.11 & 31.02 \\
(3c) oss-120b-low & 50 & 44.10 & 39.87 & 30.87 & 33.14 \\
(3d) oss-120b-med & 50 & 46.05 & 41.50 & 32.17 & 34.40 \\ \bottomrule
\end{tabular}}
\vspace{-0.25cm}
\end{table}
\noindent Here, $\textrm{Input}_{nc}$, $\textrm{Input}_{c}$, and $\textrm{Output}_{t}$ represent non-cached input, cached input, and total generated output tokens (including reasoning tokens), respectively. This configurable formulation enables principled comparisons across diverse infrastructure regimes by adjusting the caching discount ($\alpha \in \{0.1, 0.3, 0.5\}$) and the output premium ($\beta \in \{3, 5, 7\}$).

 In high-throughput deployments using vLLM, these parameters serve as proxies for system throughput: $\alpha$ models the efficiency of prefix reuse during prefill, while $\beta$ accounts for the significantly lower tokens per second (TPS) achieved during the resource-intensive auto-regressive decoding phase.
 In API-based environments, these weights represent the financial disparities of tiered pricing models. In particular, $\alpha$ reflects the substantial cost reductions (up to 90\%) offered by providers for cached context, while $\beta$ captures the significantly higher billing rates for generated tokens. By tuning these parameters, ETC provides a platform-agnostic framework to evaluate the economic and technical viability of deep search agents across both local backends and commercial APIs.
 
\begin{figure*}[t]
    \centering
    \begin{subfigure}[t]{0.28\textwidth}
        \centering
        \includegraphics[width=\textwidth]{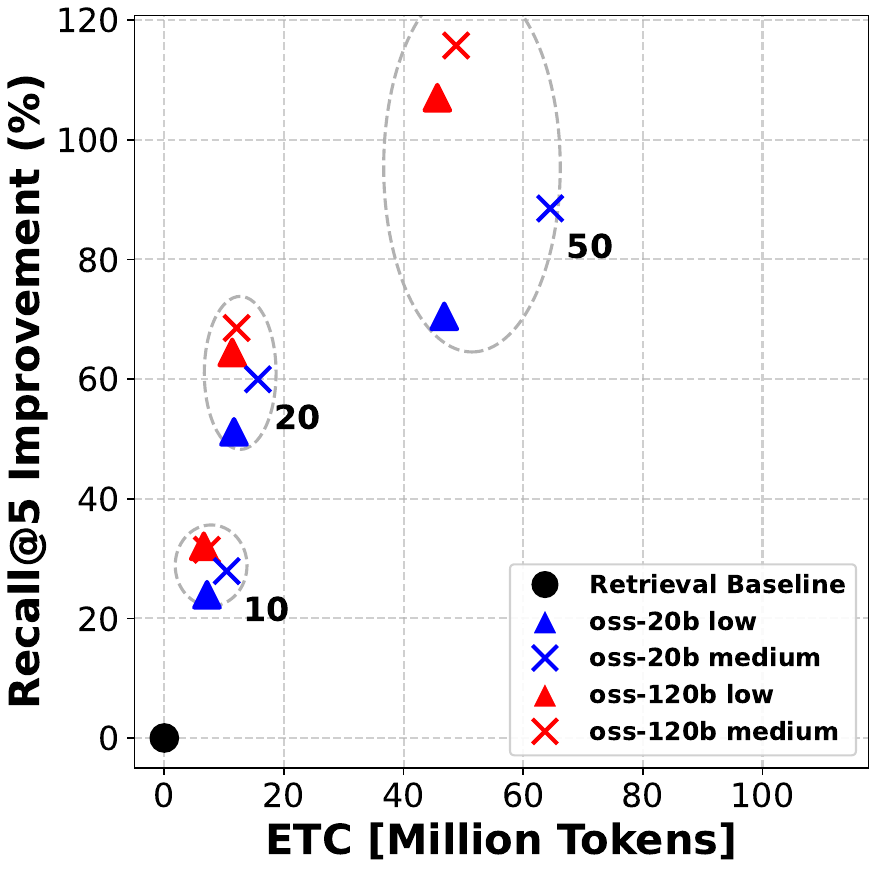}
        \caption{\textbf{$\alpha=0.1, \beta=3$}}
    \end{subfigure}
    \hfill
    \begin{subfigure}[t]{0.28\textwidth}
        \centering
        \includegraphics[width=\textwidth]{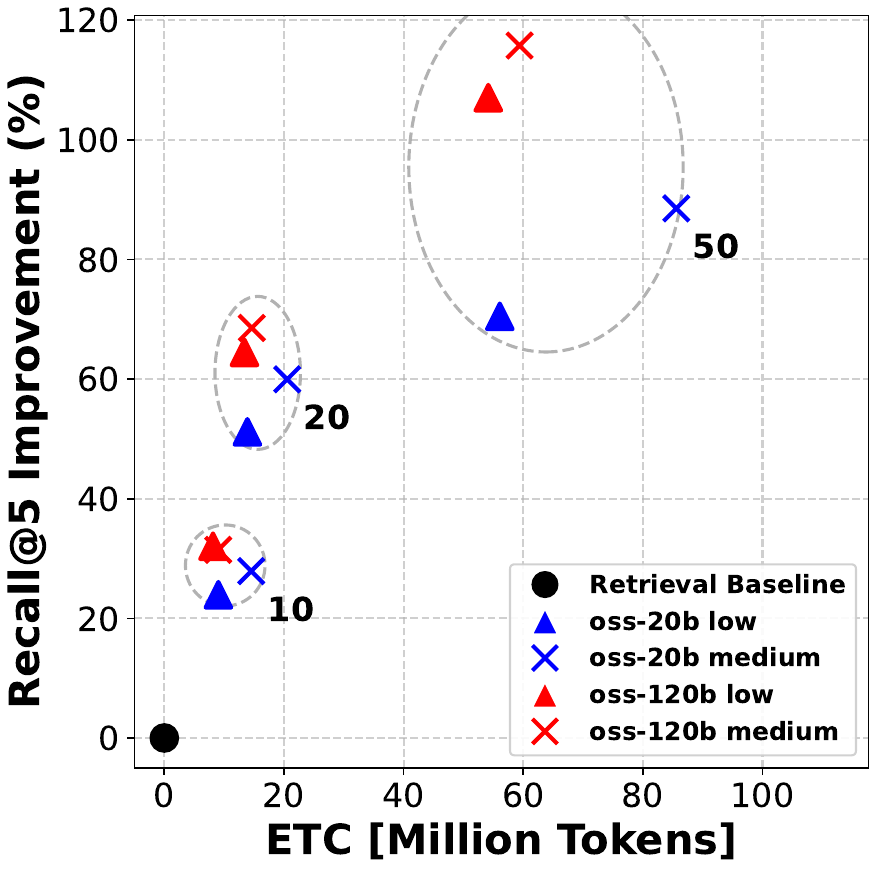}
        \caption{\textbf{$\alpha=0.1, \beta=5$}}
    \end{subfigure}
    \hfill
    \begin{subfigure}[t]{0.28\textwidth}
        \centering
        \includegraphics[width=\textwidth]{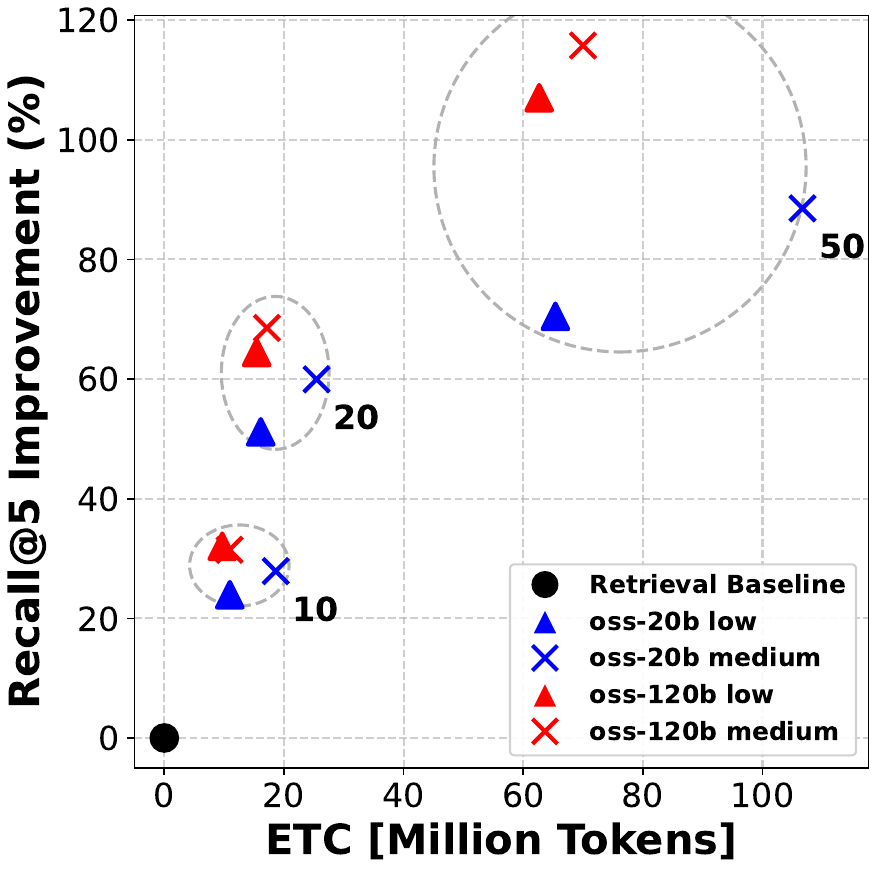}
        \caption{\textbf{$\alpha=0.1, \beta=7$}}
    \end{subfigure}
    \vspace{-2mm}
    \caption{Recall@5 gains vs.\ effective cost (per 1M tokens) for reranking depths $d \in \{10,20,50\}$ using oss-20b and oss-120b under low/medium reasoning.}
    \vspace{-2mm}
    \label{fig:recall_vs_etc}
\end{figure*}

\section{Experimental Results}
\paragraph{One-shot Reranking Effectiveness.}
We evaluate how reasoning effort and model scale influence listwise reranking performance to identify an optimal configuration prior to full deep search. Using the original query, we perform one-shot retrieval with \textit{qwen3-embedding-8b} to obtain the top 100 candidate documents. We treat evidence documents as ground-truth relevance labels. Results with gold documents---those containing the final answer---show identical trends (see \Cref{tab:retrieval_gold}).

\Cref{tab:retrieval_evidence} reports reranking results on top-$d$ candidates ($d \in \{10,20,50\}$) using oss-20b and oss-120b under low and medium reasoning budgets. Across all settings, oss-120b consistently outperforms oss-20b.
Overall, effectiveness improves monotonically within each $d$ block, with oss-120b under medium reasoning achieving the best results (e.g., NDCG@5 = 46.05 at $d=50$). Increasing reasoning budget improves ranking quality for both models, with larger gains for oss-20b, especially at higher $d$, where reranking is more difficult. For instance, at $d=50$, medium reasoning improves NDCG@5 by ~4 points for oss-20b.

However, increasing $d$ has the strongest effect. Moving from $d=10$ to $d=20$ yields ~5--7 NDCG@5 points, with an additional ~3.5--10 points from $d=20$ to $d=50$. Consequently, an oss-120b model with medium reasoning at a given $d$ can still underperform an oss-20b model with low reasoning at the next higher $d$ level.

\begin{figure*}[t]
    \centering
    \begin{subfigure}[t]{0.32\textwidth}
        \centering
        \includegraphics[width=\textwidth]{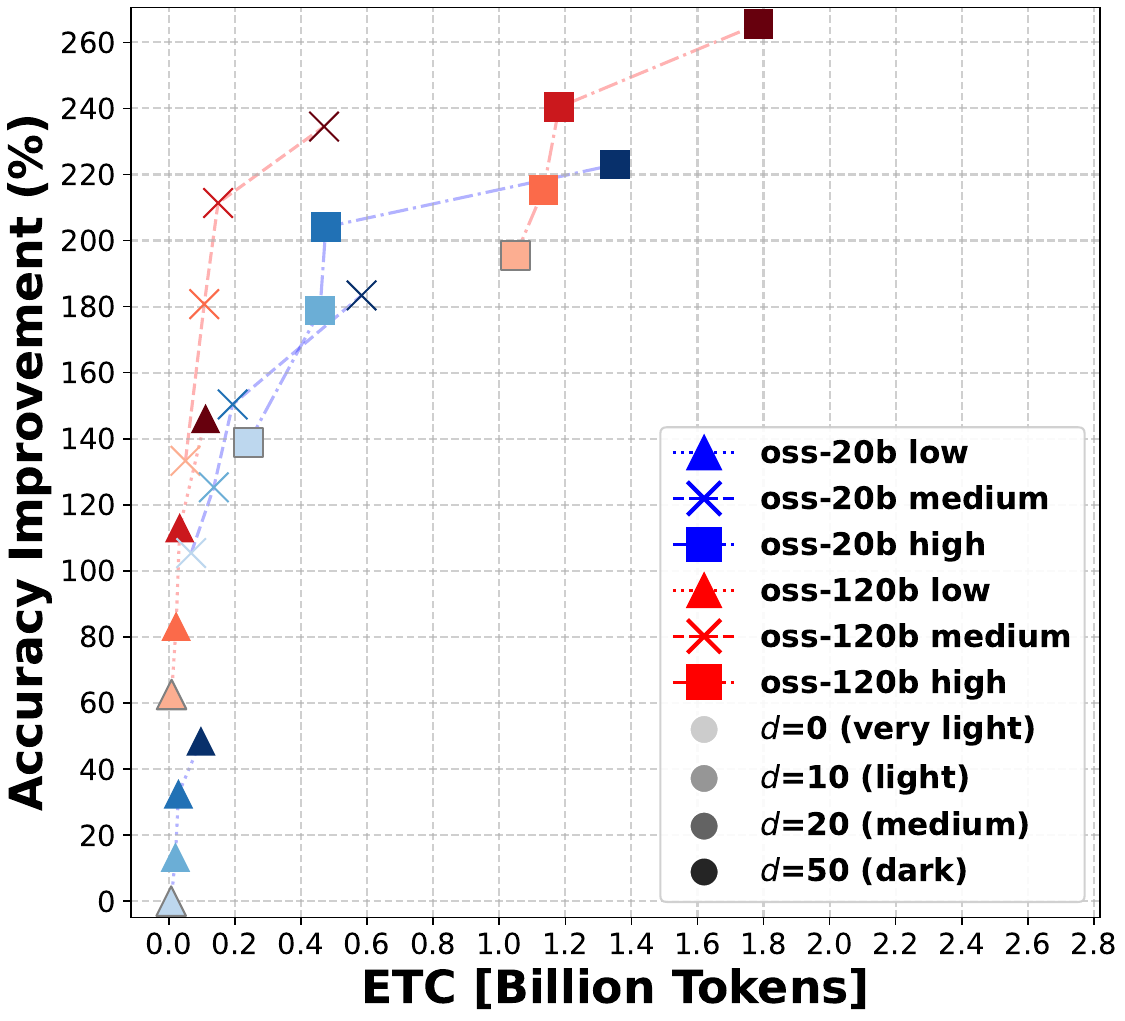}
        \caption{\textbf{$\alpha=0.1, \beta=3$}}
    \end{subfigure}
    \hfill
    \begin{subfigure}[t]{0.32\textwidth}
        \centering
        \includegraphics[width=\textwidth]{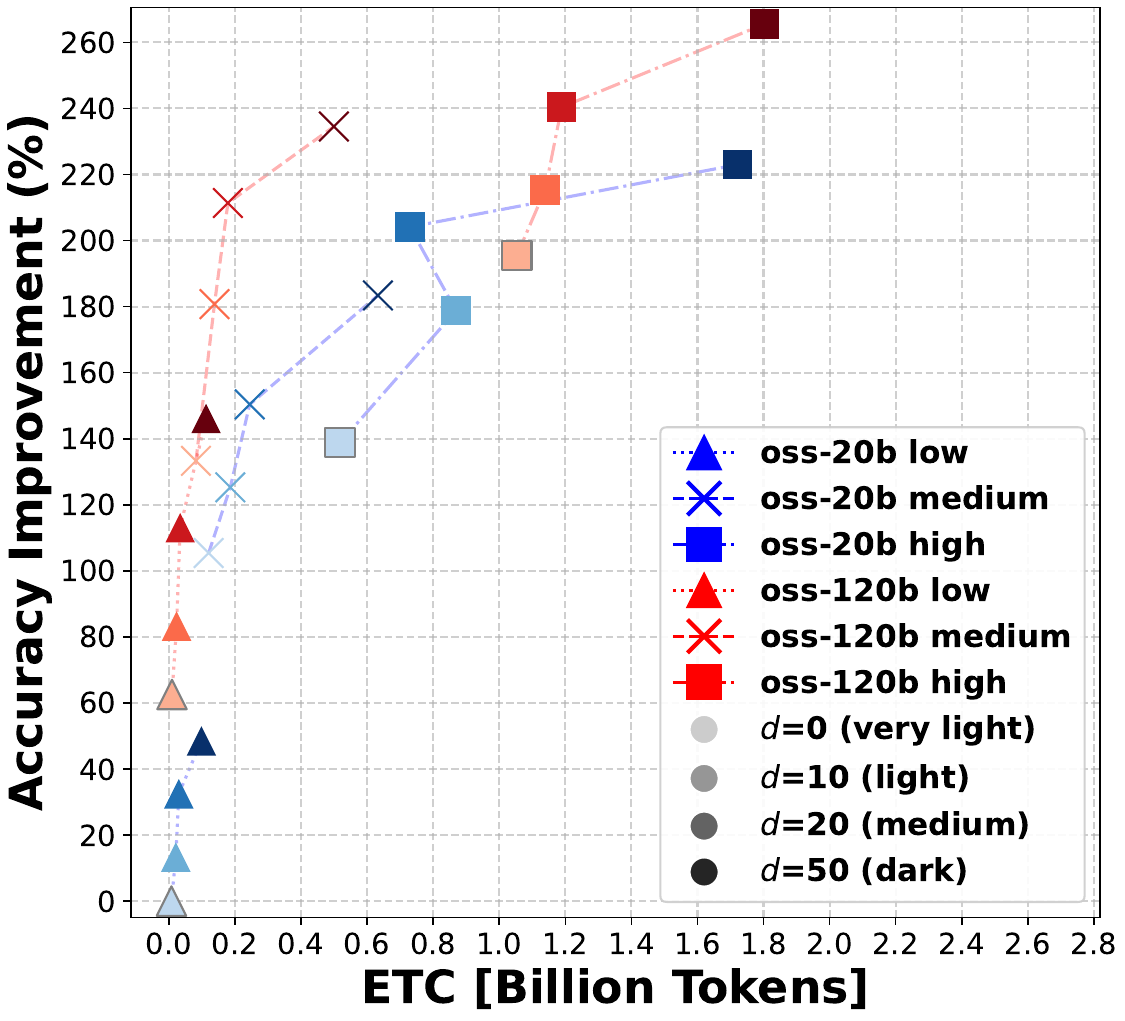}
        \caption{\textbf{$\alpha=0.3, \beta=3$}}
    \end{subfigure}
    \hfill
    \begin{subfigure}[t]{0.32\textwidth}
        \centering
        \includegraphics[width=\textwidth]{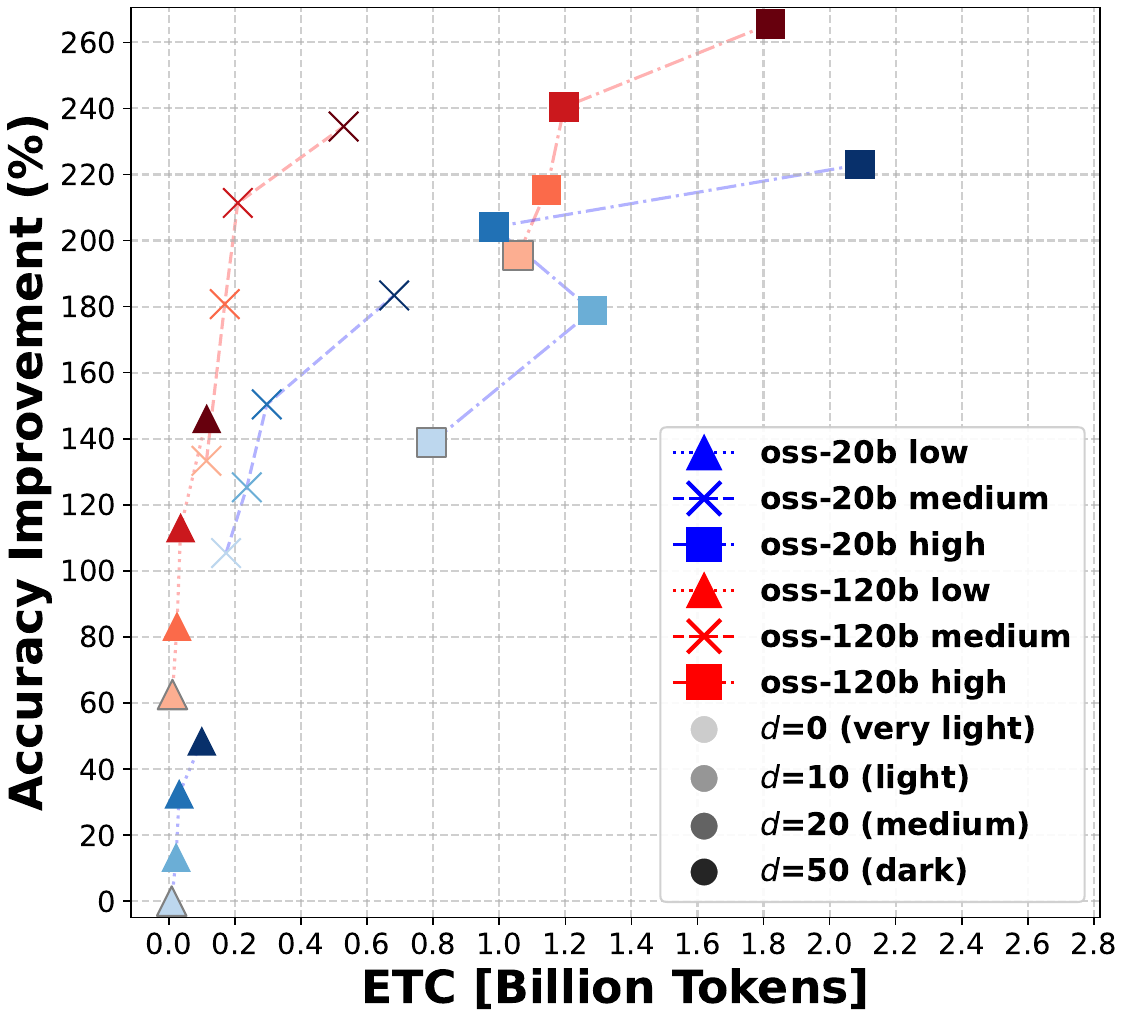}
        \caption{\textbf{$\alpha=0.5, \beta=3$}}
    \end{subfigure}
    \vspace{-3mm}\caption{Accuracy gains vs.\ effective cost (per 10M tokens) for oss-20b and oss-120b under low/medium/high reasoning with reranking depths $d \in \{0,10,20,50\}$; top-5 candidates are passed to the search agent in all cases.}
    \vspace{-4mm}
    \label{fig:accuracy_vs_etc}
\end{figure*}
\begin{table}[tbh]
\centering

\caption{End-to-end effectiveness of deep research agents under low/medium/high reasoning with reranking depths $d \in \{0,10,20,50\}$. The final top-5 candidates are passed to the agent. Accuracy (Acc.) and Calibration Error (Calib.) report means with 95\% CIs.}
\vspace{-0.1cm}
\label{tab:deep_search}
\resizebox{\columnwidth}{!}{
\begin{tabular}{lrrrrr}
\toprule
\textbf{Search Agent} & \textbf{d} & \textbf{Srs.} & \textbf{Recall} & \textbf{Acc. (CI)} & \textbf{Calib. (CI)} \\
\midrule
(0a) oss-20b-low    & 0  & 2.08  & 18.92 & 17.33 (0.16) & 36.62 (0.10) \\
(0b) oss-20b-med    & 0  & 13.78 & 44.00 & 35.59 (0.13) & 33.94 (0.05) \\
(0c) oss-20b-high   & 0  & 29.77 & 56.16 & 41.38 (0.17) & 22.81 (0.21) \\
(0d) oss-120b-low   & 0  & 2.49  & 25.37 & 28.17 (0.07) & 39.76 (0.36) \\
(0e) oss-120b-med   & 0  & 11.22 & 47.22 & 40.43 (0.13) & 40.14 (0.07) \\
(0f) oss-120b-high  & 0  & 26.37 & 61.43 & 51.20 (0.24) & 35.82 (0.19) \\
\midrule
(1a) oss-20b-low    & 10 & 2.10  & 22.17 & 19.64 (0.11) & 35.16 (0.09) \\
(1b) oss-20b-med    & 10 & 13.05 & 49.08 & 39.04 (0.18) & 31.33 (0.11) \\
(1c) oss-20b-high   & 10 & 28.66 & 63.38 & 48.31 (0.11) & 21.28 (0.16) \\
(1d) oss-120b-low   & 10 & 2.39  & 27.29 & 31.76 (0.17) & 36.92 (0.19) \\
(1e) oss-120b-med   & 10 & 10.76 & 53.91 & 48.65 (0.07) & 35.60 (0.13) \\
(1f) oss-120b-high  & 10 & 25.20 & 65.62 & 54.65 (0.39) & 31.83 (0.16) \\
\midrule
(2a) oss-20b-low    & 20 & 2.12  & 24.31 & 22.96 (0.08) & 35.51 (0.26) \\
(2b) oss-20b-med    & 20 & 13.18 & 54.98 & 43.39 (0.07) & 31.78 (0.07) \\
(2c) oss-20b-high   & 20 & 27.67 & 68.81 & 52.70 (0.17) & 19.70 (0.16) \\
(2d) oss-120b-low   & 20 & 2.38  & 33.13 & 36.92 (0.08) & 36.00 (0.07) \\
(2e) oss-120b-med   & 20 & 10.38 & 59.61 & 53.96 (0.07) & 34.77 (0.11) \\
(2f) oss-120b-high  & 20 & 24.51 & 68.58 & 58.99 (0.13) & 30.09 (0.22) \\
\midrule
(3a) oss-20b-low    & 50 & 2.09  & 28.43 & 25.73 (0.17) & 31.30 (0.20) \\
(3b) oss-20b-med    & 50 & 12.71 & 59.73 & 49.11 (0.08) & 28.40 (0.09) \\
(3c) oss-20b-high   & 50 & 27.09 & 71.29 & 55.97 (0.13) & 17.59 (0.00) \\
(3d) oss-120b-low   & 50 & 2.38  & 37.75 & 42.65 (0.11) & 32.34 (0.11) \\
(3e) oss-120b-med   & 50 & 10.16 & 65.60 & 57.97 (0.12) & 33.77 (0.10) \\
(3f) oss-120b-high  & 50 & 23.13 & 74.86 & 63.35 (0.12) & 29.60 (0.24) \\
\bottomrule
\end{tabular}
}
\vspace{-0.4cm}
\end{table}

\paragraph{Reranking Token Usage.} \Cref{fig:recall_vs_etc} shows Recall@5 improvement per additional one million effective tokens (ETC). Since per-token costs differ across model sizes, ETC is not used for cross-size comparison; instead, it is used to evaluate relative efficiency within each model family.

We focus on Recall@5 because only the top-5 reranked documents are passed to the downstream agent.
Results are reported for $\alpha=0.1$ only,
as fewer than 3\% of input tokens are cached
in this setting, and varying $\alpha$ does not noticeably impact the observed trends; full results are in~\Cref{fig:recall_vs_etc_all}.

For oss-120b, increasing reasoning effort has limited impact at $d<50$, with noticeable gains only at $d=50$. In contrast, oss-20b benefits from higher reasoning across all $d$. However, when normalized by ETC, low reasoning is the most efficient setting for oss-20b due to the increased $\beta$ cost from 3 to 5 and 7.

We therefore adopt low reasoning budgets for reranking in the full pipeline. This is motivated by the fact that agent-generated queries are typically simpler than the original one-shot query, reducing the need for higher reranking effort. A brief analysis of average per-query reasoning tokens for reranking (Appendix~\ref{app:token_count_analysis}) further supports this choice.

\paragraph{Deep Research Effectiveness.}
\Cref{tab:deep_search} reports end-to-end results for oss-20b and oss-120b under low/medium/high reasoning, with optional listwise reranking at depths $d\in\{0,10,20,50\}$ (rerankers use low reasoning). In all cases, the final top-5 candidates are passed to the search agent.

Using accuracy and recall as primary metrics, three trends emerge. First, for fixed model size and ranking depth, increasing reasoning effort consistently improves both metrics. Gains are larger for oss-20b than oss-120b (e.g., low vs.\ high reasoning within each $d$ block), and oss-20b with high reasoning is more effective than oss-120b with low reasoning, underscoring the importance of reasoning budget relative to model scale.

Second, for fixed model size and reasoning effort, increasing $d$ improves both recall and accuracy. However, under high reasoning at $d=50$, the models differ by $<3$ recall points but $\sim7$ accuracy points, suggesting that while retrieval improves similarly, oss-20b is less effective at converting these gains into correct final answers.

Finally, increasing $d$ only slightly reduces the number of search calls (up to 12\% for oss-120b-high at $d=50$ vs.\ 0), indicating that improvements primarily stem from better retrieval rather than changes in search behavior.

\paragraph{Deep Research Token Usage.}
\Cref{fig:accuracy_vs_etc} plots accuracy gains vs.\ additional ETC (per 10M tokens). 
We focus on $\beta=3$, since ETC
is dominated by input tokens in both search and
reranking, and varying the output premium does
not qualitatively change the observed trends (full results in~\Cref{fig:accuracy_vs_etc_all}).

Reranking yields strong gains for both models. Under low reasoning, increasing depth from \(d=0\) to \(d \in \{10,20,50\}\) gives large accuracy improvements at modest ETC. Under medium/high reasoning, gains persist for \(d \le 20\), but increasing from \(d=20\) to \(d=50\) shows diminishing returns.

\begin{table}[tbh]
\centering

\caption{Average latency measurements \textit{(seconds/query)} for oss-120b search agent at varying rank-$d$ and reasoning efforts.}
\vspace{-1mm}
\label{tab:latency}
\resizebox{0.35\columnwidth}{!}{
\begin{tabular}{cccc}
\toprule
& \multicolumn{3}{c}{\textbf{Reasoning Effort}} \\
\cmidrule(lr){2-4} 
\textbf{d} & \textbf{Low} & \textbf{Medium} & \textbf{High} \\
\midrule
0  &  0.8  & 9.0  & 184.7 \\
10 &  2.7  & 11.7 & 175.8 \\
20 &  4.0  & 16.3 & 168.3 \\
50 &  15.2 & 60.0 & 150.7 \\
\bottomrule
\end{tabular}
\vspace{-3mm}
}
\end{table}

For each model, medium reasoning with deeper reranking matches high reasoning with shallow/no reranking at lower ETC, indicating that allocating budget to reranking is often more cost-effective than increasing search-agent reasoning, especially for oss-120b.

\paragraph{Latency Analysis.}
\Cref{tab:latency} reports the average latency (seconds per query) for the larger oss-120b model across low, medium, and high reasoning settings. Latency is implementation-dependent; we report total wall-clock time normalized by the number of completed queries. All experiments use 32 threads, and reranking is batched via a queue (maximum size 16, maximum wait time 2000~ms, idle flush 500~ms).

As expected, latency increases with reranking depth under low and medium reasoning. In contrast, under high reasoning, latency decreases, driven by fewer search calls and concurrent utilization of the reranker and search agent. Overall, medium reasoning with moderate reranking depth provides the most favorable tradeoff: it is substantially faster than high reasoning while improving end-to-end accuracy and reducing ETC.

\begin{table}[tbh]
\centering

\caption{End-to-end effectiveness of deep research agents under low/medium/high reasoning with reranking depths $d \in \{0,10,20,50\}$ using qwen3-reranker-0.6b. The final top-5 candidates are passed to the search agent. Accuracy (Acc.) and Calibration Error (Calib.) are reported as means with 95\% CIs.}

\vspace{-0.1cm}
\label{tab:deep_search_pointwise}
\resizebox{\columnwidth}{!}{
\begin{tabular}{lrrrrr}
\toprule
\textbf{Search Agent} & \textbf{d} & \textbf{Srs.} & \textbf{Recall} & \textbf{Acc. (CI)} & \textbf{Calib. (CI)} \\
\midrule
(0a) oss-20b-low    & 0  & 2.08  & 18.92 & 17.33 (0.16) & 36.62 (0.10) \\
(0b) oss-20b-med    & 0  & 13.78 & 44.00 & 35.59 (0.13) & 33.94 (0.05) \\
(0c) oss-20b-high   & 0  & 29.77 & 56.16 & 41.38 (0.17) & 22.81 (0.21) \\
\midrule
(1a) oss-20b-low    & 10 & 2.16  & 19.70 & 18.41 (0.12) & 34.64 (0.40) \\
(1b) oss-20b-med    & 10 & 13.65 & 47.64 & 39.14 (0.22) & 30.42 (0.32) \\
(1c) oss-20b-high   & 10 & 28.56 & 59.96 & 46.44 (0.20) & 22.83 (0.05) \\
\midrule
(2a) oss-20b-low    & 20 & 2.13  & 21.22 & 19.28 (0.11) & 33.93 (0.12) \\
(2b) oss-20b-med    & 20 & 13.46 & 51.75 & 43.11 (0.07) & 31.41 (0.02) \\
(2c) oss-20b-high   & 20 & 28.59 & 63.47 & 48.38 (0.17) & 22.04 (0.38) \\
\midrule
(3a) oss-20b-low    & 50 & 2.13  & 22.79 & 23.37 (0.11) & 32.83 (0.16) \\
(3b) oss-20b-med    & 50 & 13.13 & 53.52 & 44.70 (0.18) & 29.42 (0.08) \\
(3c) oss-20b-high   & 50 & 27.67 & 66.63 & 52.91 (0.16) & 18.27 (0.24) \\
\bottomrule
\end{tabular}

}
\end{table}
\paragraph{Heterogeneous Setup.}
Previous sections assume a homogeneous setup where the same model is used for both the agent and reranker, and ETC depends only on token volume. We now consider a heterogeneous setting to generalize the efficiency--effectiveness tradeoff.

We use qwen3-reranker-0.6b as a cross-encoder (prompt in \Cref{fig:reranking-prompt-pointwise}), restricted to binary (yes/no) outputs, and rank documents using the logits of the \textit{yes} token. \Cref{tab:deep_search_pointwise} shows that even lightweight rerankers improve end-to-end accuracy and recall, though with smaller gains than the homogeneous setup (e.g., 52.91\% vs.\ 55.97\% accuracy for oss-20b-high at $d=50$, and 66.63\% vs.\ 71.29\% recall). All trends remain consistent, with performance improving as $d$ increases.
We generalize ETC as:
\begin{equation}
\textrm{ETC} = \textrm{ETC}_{agent} + \gamma \cdot \textrm{ETC}_{reranker}
\end{equation}
where $\gamma$ reflects relative inference FLOPs. For qwen3-reranker-0.6b vs.\ gpt-oss-20b, $\gamma = 0.32$ (Appendix~\ref{app:gflops}).

\begin{figure}[tbh]
    \centering
        \includegraphics[width=0.6\columnwidth]{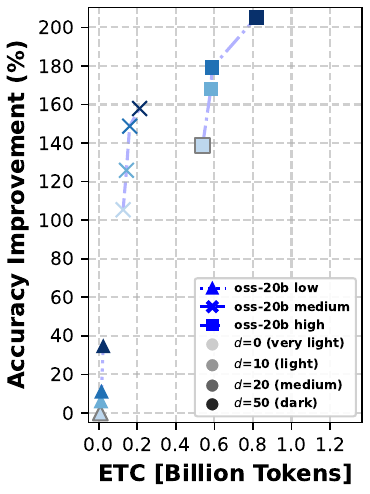}

    \vspace{-3mm}\caption{Accuracy gains vs. effective cost (per 10M tokens) with \textbf{$\alpha=0.3,\ \beta=5$} for oss-20b under low/medium/high reasoning and reranking depths $d \in \{0,10,20,50\}$ using qwen3-reranker-0.6b; top-5 candidates are passed to the search agent in all cases.}

    \label{fig:accuracy_vs_etc_pointwise}
\end{figure}

\Cref{fig:accuracy_vs_etc_pointwise} plots accuracy gains per 10M tokens. Results use median $\alpha$ and $\beta$, with similar trends across all values (\Cref{fig:accuracy_vs_etc_all_pointwise}). Since $\gamma \ll 1$, cross-encoder reranking is more cost-effective than listwise reranking. 
Additionally, listwise reranking at $d=50$ incurs input duplication due to overlapping sliding windows (each document appears $\sim$1.6$\times$ on average), making depth scaling less efficient. Thus, while listwise reranking achieves slightly higher peak accuracy, cross-encoders are preferable when efficiency is the priority.

\section{Conclusion}
We study how to allocate reasoning budget in deep research agents, focusing on listwise reranking. Reranking consistently improves retrieval and end-to-end accuracy, while increasing search-time reasoning yields diminishing returns at higher token cost. Using the effective token cost (ETC) metric, we show that moderate reranking is often more cost-effective than increasing reasoning during search.

Separately, our latency analysis shows that moderate reranking can also reduce end-to-end latency compared to high reasoning settings, providing a better efficiency--performance tradeoff.
We further show that heterogeneous setups with lightweight cross-encoder rerankers improve efficiency, offering favorable accuracy--cost tradeoffs despite slightly lower peak performance than homogeneous listwise reranking. Overall, these results identify reranking as a key lever for building efficient deep research systems.

\section*{Acknowledgments}
This research was supported in part by the Natural Sciences and Engineering Research Council (NSERC) of Canada. 
Additional funding was provided by Snowflake and the Institute of Information \& Communications Technology Planning \& Evaluation (IITP) grant funded by the Korean Government (MSIT) (No.\ RS-2024-00457882, National AI Research Lab Project).
The authors thank Zijian Chen, Xueguang Ma, and other BrowseComp-Plus authors for providing the dataset, codebase, and helpful feedback.
The authors are also grateful to Sayeh Sharify for guidance on FLOPs calculation.

\section*{Limitations}
First, we focus on the gpt-oss family; extending experiments to commercial models with controllable reasoning effort would strengthen generality.

Second, experiments use BrowseComp-Plus with a fixed, human-verified corpus rather than live web search, so results may not fully transfer to settings relying on external search APIs.

Third, reranking is implemented as a fixed top-$d$ selection. A learned relevance model that adaptively selects a variable-sized subset could better filter redundant or irrelevant context. Finally, we retain full interaction history with automatic truncation at 128k tokens; explicit compression or summarization could further reduce token usage and improve efficiency.

\bibliography{main}

\begin{thebibliography}{15}
\providecommand{\natexlab}[1]{#1}

\bibitem[{Asai et~al.(2024)Asai, Wu, Wang, Sil, and Hajishirzi}]{selfrag}
Akari Asai, Zeqiu Wu, Yizhong Wang, Avirup Sil, and Hannaneh Hajishirzi. 2024.
\newblock Self-{RAG}: Learning to retrieve, generate, and critique through self-reflection.
\newblock In \emph{Proceedings of the 12th International Conference on Learning Representations (ICLR 2024)}, Vienna, Austria.

\bibitem[{Chen et~al.(2025)Chen, Ma, Zhuang, Nie, Zou, Liu, Green, Patel, Meng, Su, Sharifymoghaddam, Li, Hong, Shi, Liu, Thakur, Zhang, Gao, Chen, and Lin}]{browsecompplus}
Zijian Chen, Xueguang Ma, Shengyao Zhuang, Ping Nie, Kai Zou, Andrew Liu, Joshua Green, Kshama Patel, Ruoxi Meng, Mingyi Su, Sahel Sharifymoghaddam, Yanxi Li, Haoran Hong, Xinyu Shi, Xuye Liu, Nandan Thakur, Crystina Zhang, Luyu Gao, Wenhu Chen, and Jimmy Lin. 2025.
\newblock {BrowseComp-Plus}: A more fair and transparent evaluation benchmark of deep-research agent.
\newblock \emph{arXiv:2508.06600}.

\bibitem[{Gim et~al.(2024)Gim, Chen, Lee, Sarda, Khandelwal, and Zhong}]{gim2024prompt}
In~Gim, Guojun Chen, Seung-seob Lee, Nikhil Sarda, Anurag Khandelwal, and Lin Zhong. 2024.
\newblock Prompt cache: Modular attention reuse for low-latency inference.
\newblock In \emph{Proceedings of the 6th Annual Conference on Machine Learning and Systems (MLSys 2024)}, pages 325--338, Santa Clara, CA, USA.

\bibitem[{Huang et~al.(2025)Huang, Chen, Zhang, Li, Zhou, Fang, Yang, Li, Shang, Xu, Hao, Shao, and Wang}]{huang2025deep}
Yuxuan Huang, Yihang Chen, Haozheng Zhang, Kang Li, Huichi Zhou, Meng Fang, Linyi Yang, Xiaoguang Li, Lifeng Shang, Songcen Xu, Jianye Hao, Kun Shao, and Jun Wang. 2025.
\newblock Deep research agents: A systematic examination and roadmap.
\newblock \emph{arXiv:2506.18096}.

\bibitem[{Lewis et~al.(2020)Lewis, Perez, Piktus, Petroni, Karpukhin, Goyal, Küttler, Lewis, tau Yih, Rocktäschel, Riedel, and Kiela}]{lewis2020retrieval}
Patrick Lewis, Ethan Perez, Aleksandra Piktus, Fabio Petroni, Vladimir Karpukhin, Naman Goyal, Heinrich Küttler, Mike Lewis, Wen tau Yih, Tim Rocktäschel, Sebastian Riedel, and Douwe Kiela. 2020.
\newblock Retrieval-augmented generation for knowledge-intensive {NLP} tasks.
\newblock In \emph{Proceedings of the 34th Annual Conference on Neural Information Processing Systems (NeurIPS 2020)}, pages 9459--9474.

\bibitem[{Nogueira and Cho(2019)}]{monobert}
Rodrigo Nogueira and Kyunghyun Cho. 2019.
\newblock Passage re-ranking with {BERT}.
\newblock \emph{arXiv:1901.04085}.

\bibitem[{Nogueira et~al.(2020)Nogueira, Jiang, Pradeep, and Lin}]{monot5}
Rodrigo Nogueira, Zhiying Jiang, Ronak Pradeep, and Jimmy Lin. 2020.
\newblock Document ranking with a pretrained sequence-to-sequence model.
\newblock In \emph{Findings of the association for computational linguistics: EMNLP 2020}, pages 708--718.

\bibitem[{Phan et~al.(2025)Phan, Gatti, Han, Li, Hu, Zhang, Zhang, Shaaban, Ling, Shi, Choi, Agrawal, Chopra, Khoja, Kim, Ren, Hausenloy, Zhang, Mazeika, Dodonov, Nguyen, Lee, Anderson, Doroshenko, Stokes, Mahmood, Pokutnyi, Iskra, Wang, Levin, Kazakov, Feng, Feng, Zhao, Yu, Gangal, Zou, Wang, Popov, Gerbicz, Galgon, Schmitt, Yeadon, Lee, Sauers, Sanchez, Giska, Roth, Riis, Utpala, Burns, Goshu, Naiya, Agu, Giboney, Cheatom, Fournier-Facio, Crowson, Finke, Cheng, Zampese, Hoerr, Nandor, Park, Gehrunger, Cai, McCarty, Garretson, Taylor, Sileo, Ren, Qazi, Li, Nam, Wydallis, Arkhipov, Shi, Bacho, Willcocks, Cao, Motwani, de~Oliveira~Santos, Veith, Vendrow, Cojoc, Zenitani, Robinson, Tang, Li, Vendrow, Fraga, Kuchkin, Maksimov, Marion, Efremov, Lynch, Liang, Mikov, Gritsevskiy, Guillod, Demir, Martinez, Pageler, Zhou, Soori, Press, Tang, Rissone, Green, Brüssel, Twayana, Dieuleveut, Imperial, Prabhu, Yang, Crispino, Rao, Zvonkine, Loiseau, Kalinin, Lukas, Manolescu, Stambaugh, Mishra, Hogg, Bosio, Coppola,
  Salazar, Jin, Sayous, Ivanov, Schwaller, Senthilkuma, Bran, Algaba, den Houte, Sypt, Verbeken, Noever, Kopylov, Myklebust, Li, Schut, Zheltonozhskii, Yuan, Lim, Stanley, Yang, Maar, Wykowski, Oller, Sahu, Ardito, Hu, Kamdoum, Jin, Vilchis, Zu, Lackner, Koppel, Sun, Antonenko, Chern, Zhao, Arsene, Cavanagh, Li, Shen, Crisostomi, Zhang, Dehghan, Ivanov, Perrella, Kaparov, Zang, Sucholutsky, Kharlamova, Orel, Poritski, Ben-David, Berger, Whitfill, Foster, Munro, Ho, Sivarajan, Hava, Kuchkin, Holmes, Rodriguez-Romero, Sommerhage, Zhang, Moat, Schneider, Kazibwe, Clarke, Kim, Dias, Fish, Elser, Kreiman, Vilchis, Klose, Anantheswaran, Zweiger, Rawal, Li, Nguyen, Daans, Heidinger, Radionov, Rozhoň, Ginis, Stump, Cohen, Poświata, Tkadlec, Goldfarb, Wang, Padlewski, Barzowski, Montgomery, Stendall, Tucker-Foltz, Stade, Rogers, Goertzen, Grabb, Shukla, Givré, Ambay, Sen, Aziz, Inlow, He, Zhang, Kaddar, Ängquist, Chen, Wang, Ramakrishnan, Thornley, Terpin, Schoelkopf, Zheng, Carmi, Brown, Zhu, Bartolo, Wheeler,
  Stehberger, Bradshaw, Heimonen, Sridhar, Akov, Sandlin, Makarychev, Tam, Hoang, Cunningham, Goryachev, Patramanis, Krause, Redenti, Aldous, Lai, Coleman, Xu, Lee, Magoulas, Zhao, Tang, Cohen, Paradise, Kirchner, Ovchynnikov, Matos, Shenoy, Wang, Nie, Sztyber-Betley, Faraboschi, Riblet, Crozier, Halasyamani, Verma, Joshi, Meril, Ma, Andréoletti, Singhal, Platnick, Nevirkovets, Basler, Ivanov, Khoury, Gustafsson, Piccardo, Mostaghimi, Chen, Singh, Khánh, Rosu, Szlyk, Brown, Narayan, Menezes, Roberts, Alley, Sun, Patel, Lamparth, Reuel, Xin, Xu, Loader, Martin, Wang, Achilleos, Preu, Korbak, Bosio, Kazemi, Chen, Bálint, Lo, Wang, Nunes, Milbauer, Bari, Wang, Ansarinejad, Sun, Durand, Elgnainy, Douville, Tordera, Balabanian, Wolff, Kvistad, Milliron, Sakor, Eron, O., Shah, Zhou, Kamalov, Abdoli, Santens, Barkan, Tee, Zhang, Tomasiello, Luca, Looi, Le, Kolt, Pan, Rodman, Drori, Fossum, Muennighoff, Jagota, Pradeep, Fan, Eicher, Chen, Thaman, Merrill, Firsching, Harris, Ciobâcă, Gross, Pandey, Gusev, Jones,
  Agnihotri, Zhelnov, Mofayezi, Piperski, Zhang, Dobarskyi, Leventov, Soroko, Duersch, Taamazyan, Ho, Ma, Held, Xian, Zebaze, Mohamed, Leser, Yuan, Yacar, Lengler, Olszewska, Fratta, Oliveira, Jackson, Zou, Chidambaram, Manik, Haffenden, Stander, Dasouqi, Shen, Golshani, Stap, Kretov, Uzhou, Zhidkovskaya, Winter, Rodriguez, Lauff, Wehr, Tang, Hossain, Phillips, Samuele, Ekström, Hammon, Patel, Farhidi, Medley, Mohammadzadeh, Peñaflor, Kassahun, Friedrich, Perez, Pyda, Sakal, Dhamane, Mirabadi, Hallman, Okutsu, Battaglia, Maghsoudimehrabani, Amit, Hulbert, Pereira, Weber, Handoko, Peristyy, Malina, Mehkary, Aly, Reidegeld, Dick, Friday, Singh, Shapourian, Kim, Costa, Gurdogan, Kumar, Ceconello, Zhuang, Park, Carroll, Tawfeek, Steinerberger, Aggarwal, Kirchhof, Dai, Kim, Ferret, Shah, Wang, Yan, Burdzy, Zhang, Franca, Pham, Loh, Robinson, Jackson, Giordano, Petersen, Cosma, Colino, White, Votava, Vinnikov, Delaney, Spelda, Stritecky, Shahid, Mourrat, Vetoshkin, Sponselee, Bacho, Yong, de~la Rosa, Cho, Li,
  Malod, Weller, Albani, Lang, Laurendeau, Kazakov, Adesanya, Portier, Hollom, Souza, Zhou, Degorre, Yalın, Obikoya, Rai, Bigi, Boscá, Shumar, Bacho, Recchia, Popescu, Shulga, Tanwie, Lux, Rank, Ni, Brooks, Yakimchyk, Huanxu, Liu, Cavalleri, Häggström, Verkama, Newbould, Gundlach, Brito-Santana, Amaro, Vajipey, Grover, Wang, Kratish, Li, Gopi, Caciolai, de~Witt, Hernández-Cámara, Rodolà, Robins, Williamson, Cheng, Raynor, Qi, Segev, Fan, Martinson, Wang, Hausknecht, Brenner, Mao, Demian, Kassani, Zhang, Avagian, Scipio, Ragoler, Tan, Sims, Plecnik, Kirtland, Bodur, Shinde, Labrador, Adoul, Zekry, Karakoc, Santos, Shamseldeen, Karim, Liakhovitskaia, Resman, Farina, Gonzalez, Maayan, Anderson, Pena, Kelley, Mariji, Pouriamanesh, Wu, Finocchio, Alarab, Cole, Ferreira, Johnson, Safdari, Dai, Arthornthurasuk, McAlister, Moyano, Pronin, Fan, Ramirez-Trinidad, Malysheva, Pottmaier, Taheri, Stepanic, Perry, Askew, Rodríguez, Minissi, Lorena, Iyer, Fasiludeen, Clark, Ducey, Piza, Somrak, Vergo, Qin, Borbás,
  Chu, Lindsey, Jallon, McInnis, Chen, Semler, Gloor, Shah, Carauleanu, Lauer, Đuc Huy, Shahrtash, Duc, Lewark, Brown, Albanie, Weber, Vaz, Clavier, Fan, e~Silva, Long, Lian, Abramovitch, Jiang, Mendoza, Islam, Gonzalez, Mavroudis, Xu, Kumar, Goswami, Bugas, Heydari, Jeanplong, Jansen, Pinto, Apronti, Galal, Ze-An, Singh, Jiang, of~Arc~Xavier, Agarwal, Berkani, Zhang, Du, de~Oliveira~Junior, Malishev, Remy, Hartman, Tarver, Mensah, Loume, Morak, Habibi, Hoback, Cai, Gimenez, Montecillo, Łucki, Campbell, Sharma, Meer, Gul, Gonzalez, Alapont, Hoover, Chhablani, Vargus, Agarwal, Jiang, Patil, Outevsky, Scaria, Maheshwari, Dendane, Shukla, Cartwright, Bogdanov, Mündler, Möller, Arnaboldi, Thaman, Siddiqi, Saxena, Gupta, Fruhauff, Sherman, Vincze, Usawasutsakorn, Ler, Radhakrishnan, Enyekwe, Salauddin, Muzhen, Maksapetyan, Rossbach, Harjadi, Bahaloohoreh, Sparrow, Sidhu, Ali, Bian, Lai, Singer, Uro, Bateman, Sayed, Menshawy, Duclosel, Bezzi, Jain, Aaron, Tiryakioglu, Siddh, Krenek, Shah, Jin, Creighton,
  Peskoff, EL-Wasif, V, Richmond, McGowan, Patwardhan, Sun, Sun, Zubić, Sala, Ebert, Kaddour, Schottdorf, Wang, Petruzella, Meiburg, Medved, ElSheikh, Hebbar, Vaquero, Yang, Poulos, Zouhar, Bogdanik, Zhang, Sanz-Ros, Anugraha, Dai, Nhu, Wang, Demircali, Jia, Zhou, Wu, He, Chandok, Sinha, Luo, Le, Noyé, Perełkiewicz, Pantidis, Qi, Purohit, Parcalabescu, Nguyen, Winata, Ponti, Li, Dhole, Park, Abbondanza, Wang, Nayak, Caetano, Wong, del Rio-Chanona, Kondor, Francois, Chalstrey, Zsambok, Hoyer, Reddish, Hauser, Rodrigo-Ginés, Datta, Shepherd, Kamphuis, Zhang, Kim, Sun, Yao, Dernoncourt, Krishna, Rismanchian, Pu, Pinto, Wang, Shridhar, Overholt, Briia, Nguyen, David, Bartomeu, Pang, Wecker, Xiong, Li, Huber, Jaeger, Maddalena, Lù, Zhang, Beger, Kon, Li, Sanker, Yin, Liang, Zhang, Agrawal, Yifei, Zhang, Cai, Sonmez, Cozianu, Li, Slen, Yu, Park, Sarti, Briański, Stolfo, Nguyen, Zhang, Perlitz, Hernandez-Orallo, Li, Shabani, Juefei-Xu, Dhingra, Zohar, Nguyen, Pondaven, Yilmaz, Zhao, Jin, Jiang, Todoran, Han,
  Kreuer, Rabern, Plassart, Maggetti, Yap, Geirhos, Kean, Wang, Mollaei, Sun, Yin, Wang, Li, Chang, Wei, Bizeul, Wang, Arrais, Mukherjee, Chamorro-Padial, Liu, Qu, Guan, Bouyamourn, Wu, Plomecka, Chen, Tang, Deng, Subramanian, Xi, Chen, Zhang, Ren, Tu, Kim, Chen, Marjanović, Ha, Luczyna, Ma, Shen, Song, Zhang, Wang, Gendron, Xiao, Smucker, Weng, Lee, Ye, Ermon, Lopez-Miguel, Knights, Gitter, Park, Wei, Chen, Pai, Elkhanany, Lin, Siedler, Fang, Mishra, Zsolnai-Fehér, Jiang, Khan, Yuan, Jain, Lin, Peterson, Wang, Malusare, Tang, Gupta, Fosin, Kang, Dworakowska, Matsumoto, Zheng, Sewuster, Villanueva, Rannev, Chernyavsky, Chen, Banik, Racz, Dong, Wang, Bashmal, Gonçalves, Hu, Bar, Bohdal, Patlan, Dhuliawala, Geirhos, Wist, Kansal, Chen, Tire, Yücel, Christof, Singla, Song, Chen, Ge, Ponkshe, Park, Shi, Ma, Mak, Lai, Moulin, Cheng, Zhu, Zhang, Patil, Jha, Men, Wu, Zhang, Vieira, Aji, Chung, Mahfoud, Hoang, Sperzel, Hao, Meding, Xu, Kostakos, Manini, Liu, Toukmaji, Paek, Yu, Demircali, Sun, Dewerpe, Qin,
  Pflugfelder, Bailey, Morris, Heilala, Rosset, Yu, Chen, Yeo, Jain, Yang, Chigurupati, Chernyavsky, Reddy, Venugopalan, Batra, Park, Tran, Maximiano, Zhang, Liang, Shiyu, Xu, Pan, Suresh, Liu, Gulati, Zhang, Turchin, Bartlett, Scotese, Cao, Wu, Karwowski, Scaramuzza, Nattanmai, McKellips, Cheraku, Suhail, Luo, Deng, Luo, Zhang, Jindel, Paek, Halevy, Baranov, Liu, Avadhanam, Zhang, Cheng, Ma, Fu, Do, Lass, Yang, Sunkari, Bharath, Ai, Leung, Agrawal, Zhou, Chen, Kalpathi, Xu, Wang, Xiao, Maung, Lee, Yang, Yue, Zhao, Yoon, Sun, Singh, Luo, Peng, Osbey, Wang, Echeazu, Yang, Wu, Patel, Kulkarni, Sundarapandiyan, Zhang, Le, Nasim, Yalam, Kasamsetty, Samal, Yang, Sun, Shah, Saha, Zhang, Nguyen, Nagumalli, Wang, Zhou, Wu, Luo, Telluri, Yue, Wang, and Hendrycks}]{phan2025humanity}
Long Phan, Alice Gatti, Ziwen Han, Nathaniel Li, Josephina Hu, Hugh Zhang, Chen Bo~Calvin Zhang, Mohamed Shaaban, John Ling, Sean Shi, Michael Choi, Anish Agrawal, Arnav Chopra, Adam Khoja, Ryan Kim, Richard Ren, Jason Hausenloy, Oliver Zhang, Mantas Mazeika, and 1093 others. 2025.
\newblock Humanity's last exam.
\newblock \emph{arXiv:2501.14249}.

\bibitem[{Ram et~al.(2023)Ram, Levine, Dalmedigos, Muhlgay, Shashua, Leyton-Brown, and Shoham}]{ram2023context}
Ori Ram, Yoav Levine, Itay Dalmedigos, Dor Muhlgay, Amnon Shashua, Kevin Leyton-Brown, and Yoav Shoham. 2023.
\newblock In-context retrieval-augmented language models.
\newblock \emph{Transactions of the Association for Computational Linguistics}, 11:1316--1331.

\bibitem[{Sharifymoghaddam et~al.(2025)Sharifymoghaddam, Pradeep, Slavescu, Nguyen, Xu, Chen, Zhang, Chen, Xian, and Lin}]{rankllm}
Sahel Sharifymoghaddam, Ronak Pradeep, Andre Slavescu, Ryan Nguyen, Andrew Xu, Zijian Chen, Yilin Zhang, Yidi Chen, Jasper Xian, and Jimmy Lin. 2025.
\newblock Rank{LLM}: A {P}ython package for reranking with {LLMs}.
\newblock In \emph{Proceedings of the 48th International ACM SIGIR Conference on Research and Development in Information Retrieval (SIGIR 2025)}, pages 3681--3690, Padua, Italy.

\bibitem[{Sun et~al.(2023)Sun, Yan, Ma, Wang, Ren, Chen, Yin, and Ren}]{rankgpt}
Weiwei Sun, Lingyong Yan, Xinyu Ma, Shuaiqiang Wang, Pengjie Ren, Zhumin Chen, Dawei Yin, and Zhaochun Ren. 2023.
\newblock Is {C}hat{GPT} good at search? investigating large language models as re-ranking agents.
\newblock In \emph{Proceedings of the 2023 Conference on Empirical Methods in Natural Language Processing (EMNLP 2023)}, pages 14918--14937, Singapore.

\bibitem[{Wang et~al.(2025)Wang, Liu, DAI, Zeng, Zhang, Wu, Luo, Li, Tang, He, and Wang}]{agenttts}
Fali Wang, Hui Liu, Zongyu DAI, Jingying Zeng, Zhiwei Zhang, Zongyu Wu, Chen Luo, Zhen Li, Xianfeng Tang, Qi~He, and Suhang Wang. 2025.
\newblock Agent{TTS}: Large language model agent for test-time compute-optimal scaling strategy in complex tasks.
\newblock In \emph{Proceedings of the 39th Annual Conference on Neural Information Processing Systems (NeurIPS 2025)}, San Diego, USA.

\bibitem[{Xi et~al.(2025)Xi, Lin, Xiao, Zhou, Shan, Gao, Zhu, Liu, Yu, and Zhang}]{zhang2025survey}
Yunjia Xi, Jianghao Lin, Yongzhao Xiao, Zheli Zhou, Rong Shan, Te~Gao, Jiachen Zhu, Weiwen Liu, Yong Yu, and Weinan Zhang. 2025.
\newblock A survey of {LLM}-based deep search agents: Paradigm, optimization, and challenges.
\newblock \emph{arXiv:2508.05668}.

\bibitem[{Xiao et~al.(2025)Xiao, Gao, Peng, and Xiong}]{agentdiet}
Yuan-An Xiao, Pengfei Gao, Chao Peng, and Yingfei Xiong. 2025.
\newblock Improving the efficiency of {LLM} agent systems through trajectory reduction.
\newblock \emph{arXiv:2509.23586}.

\bibitem[{Zhang et~al.(2025)Zhang, Li, Long, Zhang, Lin, Yang, Xie, Yang, Liu, Lin, Huang, and Zhou}]{qwen3}
Yanzhao Zhang, Mingxin Li, Dingkun Long, Xin Zhang, Huan Lin, Baosong Yang, Pengjun Xie, An~Yang, Dayiheng Liu, Junyang Lin, Fei Huang, and Jingren Zhou. 2025.
\newblock Qwen3 embedding: Advancing text embedding and reranking through foundation models.
\newblock \emph{arXiv:2506.05176}.

\end{thebibliography}

\clearpage
\appendix
\begin{table}[tbh]
\caption{Hardware configuration for (a) reranking-only and (b) end-to-end deep research experiments.}
\label{tab:hardware_all}
\centering
\begin{minipage}{0.33\textwidth}
\resizebox{\columnwidth}{!}{
\begin{tabular}{l l l  c}
\toprule
\multicolumn{4}{c}{\textbf{Reranking-only Experiments}} \\
\midrule
Model & Reasoning & GPU Type & \#GPUs \\
\midrule
oss-20b & Low, Medium &  H200 & 1 \\
oss-120b & Low, Medium &  H200 & 1 \\
\bottomrule
\end{tabular}
}
\caption*{(a)}
\end{minipage}
\begin{minipage}{0.43\textwidth}
\resizebox{\columnwidth}{!}{
\begin{tabular}{l l l l c}
\toprule
\multicolumn{5}{c}{\textbf{End-to-end Deep Research Experiments}} \\
\midrule
Model & Reasoning & Reranking & GPU Type & \#GPUs \\
\midrule
oss-20b & low, med, high & N/A ($d=0$) & H200 & 1 \\
oss-20b & low, med, high & listwise ($d>0$) & H200 & 2 \\
oss-120b & low, med, high & N/A ($d=0$) & H200 & 1 \\
oss-120b & low, med, high & listwise ($d>0$) & H200 & 2 \\
\midrule
oss-20b & low, med, high & cross-encoder ($d>0$) & BW 6000 & 2 \\
\bottomrule
\end{tabular}
}
\caption*{(b)}
\vspace{-0.2cm}
\end{minipage}

\end{table}
\vspace{-0.3cm}
\section{Hardware Configuration}
\noindent \Cref{tab:hardware_all} summarizes the hardware configuration used across all experiments. In end-to-end runs, reranking—when enabled—is performed on a dedicated GPU to prevent interference with KV-cache utilization and overall throughput. When reranking is active, the retriever is co-located with the reranking LLM; otherwise, it is co-located with the search agent. Unless otherwise specified, all experiments use NVIDIA H200 NVL (141\,GB) GPUs; the heterogeneous setup uses NVIDIA RTX PRO 6000 Blackwell (96\,GB).

\section{LLM Prompts}
\Cref{fig:search-prompt,fig:reranking-prompt,fig:reranking-prompt-pointwise,fig:judging-prompt} show the prompt templates for deep search, listwise reranking, cross-encoder reranking, and LLM-as-a-judge evaluation, respectively.
\label{sec:appendix}

\section{Complete Reranking Results}
\label{app:complete_reranking_results}

\Cref{tab:retrieval_gold} reports reranking effectiveness under different model sizes and reasoning budgets using gold documents for query relevance assessment. Consistent with the results obtained with evidence documents, both model families benefit from increasing the number of reranked candidates and using a medium reasoning budget. The largest gains are observed for the oss-120b model with medium reasoning when reranking the top 50 candidates.

\begin{figure*}[!tbh]
\tiny
\centering
\begin{minipage}{\textwidth}
\begin{minted}[fontsize=\tiny, frame=lines, frame=single,linenos=false,breaklines,breaksymbol=,escapeinside=||,bgcolor=LightGray]{text}
"""You are a deep research agent. You need to answer the given question by interacting with a search engine, using the search tool provided. Please perform reasoning and use the tool step by step, in an interleaved manner. You may use the search tool multiple times.

Question: {Question}

Your response should be in the following format:
Explanation: {{your explanation for your final answer. For this explanation section only, you should cite your evidence documents inline by enclosing their docids in square brackets [] at the end of sentences. For example, [20].}}
Exact Answer: {{your succinct, final answer}}
Confidence: {{your confidence score between 0% and 100% for your answer}}"""
\end{minted}
\end{minipage}
\vspace*{-0.6cm}
\caption{Inference prompt for deep research agents.}
\label{fig:search-prompt}
\end{figure*}
\begin{figure*}[!tbh]
\tiny
\centering
\begin{minipage}{\textwidth}
\begin{minted}[fontsize=\tiny, frame=lines, frame=single,linenos=false,breaklines,breaksymbol=,escapeinside=||,bgcolor=LightGray]{text}
"""You are RankLLM, an intelligent assistant that can rank passages based on their relevance to the query. Given a query and a passage list, you first think about the reasoning process in mind and then provide the answer (i.e., the reranked passage list). Do not include any other text in your response.

I will provide you with {num} passages, each indicated by a numerical identifier []. Rank the passages based on their relevance to the search query: {query}.

[1] {candidate}
...

Search Query: {query}.
Rank the {num} passages above based on their relevance to the search query. All passages should be included and listed using identifiers, in descending order of relevance. The format of the answer should be [] > [], e.g., [2] > [1]."""
\end{minted}
\end{minipage}
\vspace*{-0.6cm}
\caption{Inference prompt for listwise reranking.}
\label{fig:reranking-prompt}
\end{figure*}
\begin{figure*}[!tbh]
\tiny
\centering
\begin{minipage}{\textwidth}
\begin{minted}[fontsize=\tiny, frame=lines, frame=single,linenos=false,breaklines,breaksymbol=,escapeinside=||,bgcolor=LightGray]{text}
"""Judge whether the Document meets the requirements based on the Query and the Instruct provided. Note that the answer can only be "yes" or "no".

<Instruct>: Given a web search query, retrieve relevant passages that answer the query.

<Query>: {query}

<Document>: {doc}"""
\end{minted}
\end{minipage}
\vspace*{-0.6cm}
\caption{Inference prompt for reranking with cross-encoders.}

\label{fig:reranking-prompt-pointwise}
\end{figure*}
\begin{figure*}[!tbh]
\tiny
\centering
\begin{minipage}{\textwidth}
\begin{minted}[fontsize=\tiny, frame=lines, frame=single,linenos=false,breaklines,breaksymbol=,escapeinside=||,bgcolor=LightGray]{text}
"""Judge whether the following [response] to [question] is correct or not based on the precise and unambiguous [correct_answer] below.

[question]: {question}

[response]: {response}

[correct_answer]: {correct_answer}

Your judgement must be in the format and criteria specified below:

extracted_final_answer: The final exact answer extracted from the [response]. 

[correct_answer]: Repeat the [correct_answer] given above.

reasoning: Explain why the extracted_final_answer is correct or incorrect based on [correct_answer], in the context of this [question]. You should judge whether the extracted_final_answer is semantically equivalent to [correct_answer], allowing the extracted_final_answer to be string variations of [correct_answer]. You should also allow the extracted_final_answer to be more precise or verbose than [correct_answer], as long as its additional details are correct. Do not comment on any background to the problem, do not attempt to solve the problem, do not argue for any answer different than [correct_answer], focus only on whether the answers are semantically equivalent.

correct: Answer 'yes' if extracted_final_answer matches the [correct_answer] given above, or is within a small margin of error for numerical problems. Answer 'no' otherwise, i.e. if there if there is any inconsistency, ambiguity, non-equivalency, or if the extracted answer is incorrect.

confidence: The extracted confidence score between 0|\%| and 100|\%| from [response]. Put 100 if there is no confidence score available."""
\end{minted}
\end{minipage}
\vspace*{-0.6cm}
\caption{Inference prompt for evaluating end-to-end answer correctness.}
\vspace*{-0.2cm}
\label{fig:judging-prompt}
\end{figure*}

\section{Token Count Analysis}
\label{app:token_count_analysis}
We analyze reranker token usage across query types and reasoning budgets. As shown in \Cref{tab:avg-reasoning-tokens-for-ranking}, reranking queries generated by the search agent consistently require fewer reasoning tokens than the original one-shot query across all settings.
\begin{table}[tbh]
\centering
\caption{Reranking effectiveness in a one-shot setting: the full question is used as the query, with qwen3-embedding-8b retrieving 100 candidates. The top $d \in \{10,20,50\}$ are reranked using oss-20b and oss-120b listwise rerankers under low/medium reasoning. Relevance is defined using gold documents.}
\vspace{-0.2cm}
\label{tab:retrieval_gold}
\resizebox{\columnwidth}{!}{
\begin{tabular}{@{}lrrrrr@{}}
\toprule
\textbf{Retriever/Reranker} & \textbf{d} & \textbf{NDCG@5} & \textbf{NDCG@10} & \textbf{Recall@5} & \textbf{Recall@10} \\ \midrule
(0) qwen3-emb-8b & 0 & 16.88 & 19.47 & 19.03 & 26.06 \\ \midrule
(1a) oss-20b-low & 10 & 24.28 & 24.85 & 23.37 & 26.06 \\
(1b) oss-20b-med & 10 & 25.28 & 25.53 & 23.97 & 26.06 \\
(1c) oss-120b-low & 10 & 26.21 & 26.17 & 24.67 & 26.06 \\
(1d) oss-120b-med & 10 & 26.56 & 26.61 & 24.40 & 26.06 \\ \midrule
(2a) oss-20b-low & 20 & 27.57 & 28.11 & 27.53 & 30.54 \\
(2b) oss-20b-med & 20 & 30.18 & 30.48 & 29.32 & 31.87 \\
(2c) oss-120b-low & 20 & 31.31 & 31.33 & 30.08 & 31.97 \\
(2d) oss-120b-med & 20 & 31.96 & 31.80 & 30.90 & 32.37 \\ \midrule
(3a) oss-20b-low & 50 & 31.23 & 31.77 & 31.29 & 34.28 \\
(3b) oss-20b-med & 50 & 34.59 & 35.03 & 34.61 & 37.41 \\
(3c) oss-120b-low & 50 & 37.90 & 37.82 & 37.26 & 39.14 \\
(3d) oss-120b-med & 50 & 39.68 & 39.54 & 38.52 & 40.33 \\ \bottomrule
\end{tabular}}
\vspace{-0.3cm}
\end{table} 
Moreover, higher reasoning effort during query generation further reduces reranking token consumption. For example, for oss-120b at $d=50$, the original query requires 4809.54 tokens, whereas high-effort agent-generated queries reduce this to 3764.48 tokens. This inverse relationship supports the observation that agent-refined queries are easier to rerank, justifying lower reranking budgets in the main experiments.

\setlength{\tabcolsep}{7pt}

\begin{table*}[tbh]
\centering
\caption{End-to-end token usage ($10^6$) for deep search and reranking. The total token usage is decomposed into Input, Cache, Output, and Reasoning components. Cache refers to the portion of input tokens served from cache, and Reasoning refers to the subset of output tokens used for reasoning.}
\vspace{-0.1cm}
\label{tab:deep_search_tokens}

\setlength{\tabcolsep}{5pt}

\resizebox{0.65\textwidth}{!}{
\begin{tabular}{l c
r r r r r
r r r r r}
\toprule

& &
\multicolumn{5}{c}{\textbf{Search Token Usage ($10^6$)}} &
\multicolumn{5}{c}{\textbf{Ranking Token Usage ($10^6$)}} \\
\cmidrule(lr){3-7} \cmidrule(lr){8-12}

\textbf{Search Agent} & \textbf{d}
& \textbf{Inp.} & \textbf{Cache} & \textbf{Out.} & \textbf{Reas.} & \textbf{Total}
& \textbf{Inp.} & \textbf{Cache} & \textbf{Out.} & \textbf{Reas.} & \textbf{Total} \\

\midrule

(0a) oss-20b-low   & 0  & 9.17   & 4.31   & 0.38  & 0.23  & 9.55   & -- & -- & -- & -- & -- \\
(0b) oss-20b-med   & 0  & 294.99 & 263.93 & 3.24  & 2.85  & 298.24 & -- & -- & -- & -- & -- \\
(0c) oss-20b-high  & 0  & 1456.70 & 1385.89 & 10.42 & 9.74  & 1467.12 & -- & -- & -- & -- & -- \\
(0d) oss-120b-low  & 0  & 12.05  & 6.29   & 0.46  & 0.27  & 12.51  & -- & -- & -- & -- & -- \\
(0e) oss-120b-med  & 0  & 186.30 & 158.96 & 2.17  & 1.78  & 188.47 & -- & -- & -- & -- & -- \\
(0f) oss-120b-high & 0  & 1044.14 & 19.15 & 7.41  & 6.69  & 1051.55 & -- & -- & -- & -- & -- \\

\midrule

(1a) oss-20b-low   & 10 & 9.33   & 4.43   & 0.37  & 0.22  & 9.70   & 8.78   & 0.81  & 1.66  & 1.58  & 10.43 \\
(1b) oss-20b-med   & 10 & 273.95 & 244.36 & 2.94  & 2.56  & 276.89 & 54.02  & 5.48  & 7.89  & 7.37  & 61.91 \\
(1c) oss-20b-high  & 10 & 2116.71 & 2053.07 & 9.72  & 9.04  & 2126.44 & 117.82 & 8.27  & 16.18 & 15.05 & 134.01 \\
(1d) oss-120b-low  & 10 & 11.28  & 5.73   & 0.45  & 0.26  & 11.73  & 9.99   & 0.94  & 1.63  & 1.54  & 11.62 \\
(1e) oss-120b-med  & 10 & 176.70 & 150.25 & 2.05  & 1.66  & 178.74 & 44.49  & 4.86  & 6.30  & 5.87  & 50.79 \\
(1f) oss-120b-high & 10 & 988.14 & 16.98  & 6.96  & 6.28  & 995.11 & 103.83 & 7.65  & 14.39 & 13.37 & 118.22 \\

\midrule

(2a) oss-20b-low   & 20 & 9.42   & 4.48   & 0.38  & 0.23  & 9.80   & 17.33  & 1.22  & 1.94  & 1.80  & 19.27 \\
(2b) oss-20b-med   & 20 & 280.11 & 250.30 & 2.94  & 2.55  & 283.05 & 106.50 & 7.95  & 9.89  & 8.97  & 116.39 \\
(2c) oss-20b-high  & 20 & 1312.93 & 1249.37 & 9.23  & 8.57  & 1322.16 & 221.04 & 22.11 & 19.17 & 17.23 & 240.21 \\
(2d) oss-120b-low  & 20 & 11.09  & 5.59   & 0.45  & 0.25  & 11.54  & 19.38  & 1.35  & 2.27  & 2.10  & 21.64 \\
(2e) oss-120b-med  & 20 & 166.83 & 143.10 & 1.97  & 1.58  & 168.79 & 83.82  & 6.12  & 8.76  & 8.00  & 92.58 \\
(2f) oss-120b-high & 20 & 938.71 & 21.39  & 6.80  & 6.11  & 945.51 & 196.08 & 16.98 & 20.19 & 18.40 & 216.27 \\

\midrule

(3a) oss-20b-low   & 50 & 9.38   & 4.43   & 0.37  & 0.22  & 9.75   & 68.25  & 2.20  & 7.96  & 7.43  & 76.22 \\
(3b) oss-20b-med   & 50 & 261.89 & 232.91 & 2.85  & 2.47  & 264.74 & 411.93 & 13.57 & 40.87 & 37.35 & 452.80 \\
(3c) oss-20b-high  & 50 & 1875.47 & 1814.56 & 9.12  & 8.47  & 1884.60 & 874.06 & 40.73 & 80.98 & 73.34 & 955.04 \\
(3d) oss-120b-low  & 50 & 11.39  & 5.88   & 0.45  & 0.25  & 11.84  & 77.85  & 2.42  & 9.16  & 8.48  & 87.01 \\
(3e) oss-120b-med  & 50 & 161.82 & 137.93 & 1.92  & 1.54  & 163.74 & 329.60 & 10.22 & 35.21 & 32.24 & 364.81 \\
(3f) oss-120b-high & 50 & 861.83 & 63.46  & 6.47  & 5.80  & 868.30 & 746.78 & 25.34 & 79.04 & 72.26 & 825.81 \\

\bottomrule
\end{tabular}
}
\vspace{-0.2cm}
\end{table*}

\Cref{tab:retrieval_token_stats} shows the raw token counts for reranking experiments. Input token counts are identical for $d \leq 20$ due to identical reranking prompts, while slight variations arise at $d=50$ due to the sliding-window procedure. In all cases, candidate document content dominates the prompt, resulting in most input tokens being non-cached, and the majority of output tokens corresponding to reasoning rather than structured output.

\Cref{tab:deep_search_tokens} reports raw token counts for end-to-end deep search. When $d=0$, no reranking tokens are used. Input tokens dominate overall usage for both search and reranking; however, most search input tokens are cached due to multi-turn history, while reranking inputs are largely non-cached. An exception is oss-120b under high reasoning, where cached search inputs drop significantly as the 128k context limit is reached and automatic truncation is applied.

Finally, \Cref{tab:deep_search_tokens_pointwise} reports raw token counts for the heterogeneous setup with qwen3-reranker-0.6b as a cross-encoder reranker. Comparing corresponding rows in \Cref{tab:deep_search_tokens,tab:deep_search_tokens_pointwise} shows that for $d<50$, pointwise reranking uses more input tokens: although its prompt is shorter, inference is repeated per document, whereas listwise reranking applies a longer prompt once per query. At $d=50$, however, listwise reranking incurs additional input overhead due to overlapping sliding windows, where each document appears $\sim$1.6$\times$ on average (window size 20, stride 10).
\begin{table}[ht]
\centering
\caption{Average reasoning tokens per query for original vs. agent-generated queries across low, medium, and high reasoning efforts at ranking depths $d \in \{10, 20, 50\}$.}
\label{tab:avg-reasoning-tokens-for-ranking}
\resizebox{0.75\columnwidth}{!}{
\begin{tabular}{lrrrrr}
\toprule
\textbf{d} & \textbf{Original} & \multicolumn{4}{c}{\textbf{Agent generated}} \\
\cline{3-6}
& & \textbf{Low} & \textbf{Medium} & \textbf{High} & \textbf{All} \\
\midrule
\multicolumn{6}{c}{\textit{gpt-oss-20b}} \\
10 & 1108.08 & 906.37 & 680.77 & 634.36 & 661.28 \\
20 & 1283.86 & 1021.99 & 819.93 & 756.95 & 789.54 \\
50 & 5348.67 & 4282.76 & 3539.33 & 3261.61 & 3396.84 \\
\midrule
\multicolumn{6}{c}{\textit{gpt-oss-120b}} \\
10 & 882.82 & 773.43 & 657.46 & 640.10 & 653.29 \\
20 & 1149.42 & 1061.56 & 929.74 & 909.13 & 924.65 \\
50 & 4809.54 & 4299.48 & 3822.99 & 3764.48 & 3816.80 \\
\bottomrule
\end{tabular}
}

\end{table}
\begin{table*}[tbh]
\centering
\caption{End-to-end token usage ($10^6$) for deep search and reranking. The total token usage is decomposed into Input, Cache, Output, and Reasoning components. Cache refers to the portion of input tokens served from cache, and Reasoning refers to the subset of output tokens used for reasoning. All experiments utilize qwen3-reranker-0.6b as the reranker within the pipeline.}
\label{tab:deep_search_tokens_pointwise}
\vspace{-0.1cm}
\setlength{\tabcolsep}{5pt}

\resizebox{0.65\textwidth}{!}{
\begin{tabular}{l c
r r r r r
r r r r r}
\toprule

& &
\multicolumn{5}{c}{\textbf{Search Token Usage ($10^6$)}} &
\multicolumn{5}{c}{\textbf{Ranking Token Usage ($10^6$)}} \\
\cmidrule(lr){3-7} \cmidrule(lr){8-12}

\textbf{Search Agent} & \textbf{d}
& \textbf{Inp.} & \textbf{Cache} & \textbf{Out.} & \textbf{Reas.} & \textbf{Total}
& \textbf{Inp.} & \textbf{Cache} & \textbf{Out.} & \textbf{Reas.} & \textbf{Total} \\

\midrule

(0a) oss-20b-low   & 0  & 9.17   & 4.31   & 0.38  & 0.23  & 9.55   & -- & -- & -- & -- & -- \\
(0b) oss-20b-med   & 0  & 294.99 & 263.93 & 3.24  & 2.85  & 298.24 & -- & -- & -- & -- & -- \\
(0c) oss-20b-high  & 0  & 1456.70 & 1385.89 & 10.42 & 9.74  & 1467.12 & -- & -- & -- & -- & -- \\

\midrule

(1a) oss-20b-low   & 10 & 9.83   & 4.80   & 0.38  & 0.23  & 10.21  & 10.40 & 1.41 & 0.02 & 0.00 & 10.42 \\
(1b) oss-20b-med   & 10 & 290.62 & 259.30 & 3.10  & 2.71  & 293.72 & 64.44 & 8.11 & 0.11 & 0.00 & 64.55 \\
(1c) oss-20b-high  & 10 & 1493.98 & 1431.00 & 9.99  & 9.32  & 1503.98 & 133.91 & 17.04 & 0.24 & 0.00 & 134.15 \\

\midrule

(2a) oss-20b-low   & 20 & 9.66   & 4.70   & 0.39  & 0.24  & 10.05  & 20.61 & 2.86 & 0.04 & 0.00 & 20.65 \\
(2b) oss-20b-med   & 20 & 286.36 & 255.72 & 3.10  & 2.70  & 289.46 & 127.25 & 16.08 & 0.22 & 0.00 & 127.47 \\
(2c) oss-20b-high  & 20 & 1382.45 & 1318.65 & 9.91  & 9.25  & 1392.36 & 268.66 & 33.86 & 0.47 & 0.00 & 269.13 \\

\midrule

(3a) oss-20b-low   & 50 & 9.58   & 4.63   & 0.38  & 0.23  & 9.96   & 51.24 & 7.17 & 0.09 & 0.00 & 51.33 \\
(3b) oss-20b-med   & 50 & 281.67 & 251.82 & 3.02  & 2.63  & 284.69 & 310.16 & 39.20 & 0.54 & 0.00 & 310.70 \\
(3c) oss-20b-high  & 50 & 1790.27 & 1729.16 & 9.37  & 8.71  & 1799.64 & 649.84 & 81.11 & 1.15 & 0.00 & 650.99 \\

\bottomrule
\end{tabular}
}
\vspace{-0.2cm}
\end{table*}

\section{Complete ETC Analysis}
\begin{figure*}[t]
    \centering
    \begin{subfigure}[t]{0.32\textwidth}
        \centering
        \includegraphics[width=\textwidth]{figures/etc_recall_5_a0.1_b3.pdf}
        \caption{\textbf{$\alpha=0.1, \beta=3$}}
    \end{subfigure}
    \hfill
    \begin{subfigure}[t]{0.32\textwidth}
        \centering
        \includegraphics[width=\textwidth]{figures/etc_recall_5_a0.1_b5.pdf}
        \caption{\textbf{$\alpha=0.1, \beta=5$}}
    \end{subfigure}
    \hfill
    \begin{subfigure}[t]{0.32\textwidth}
        \centering
        \includegraphics[width=\textwidth]{figures/etc_recall_5_a0.1_b7.pdf}
        \caption{\textbf{$\alpha=0.1, \beta=7$}}
    \end{subfigure}
    \vspace{2mm}
    \begin{subfigure}[t]{0.32\textwidth}
        \centering
        \includegraphics[width=\textwidth]{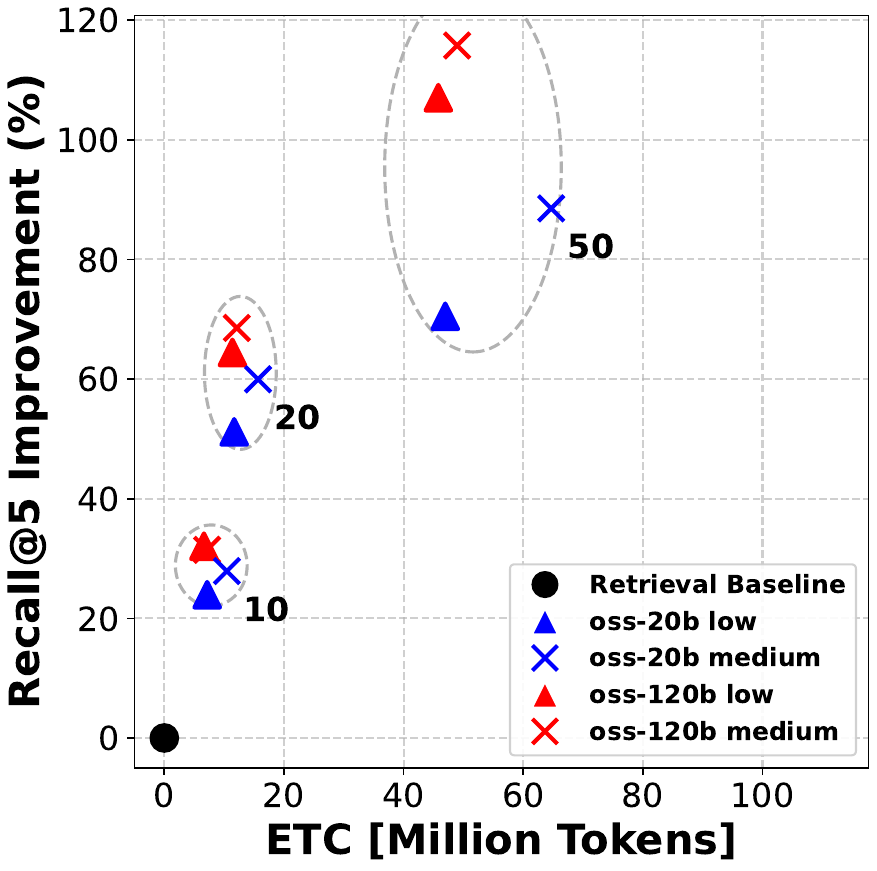}
        \caption{\textbf{$\alpha=0.3, \beta=3$}}
    \end{subfigure}
    \hfill
    \begin{subfigure}[t]{0.32\textwidth}
        \centering
        \includegraphics[width=\textwidth]{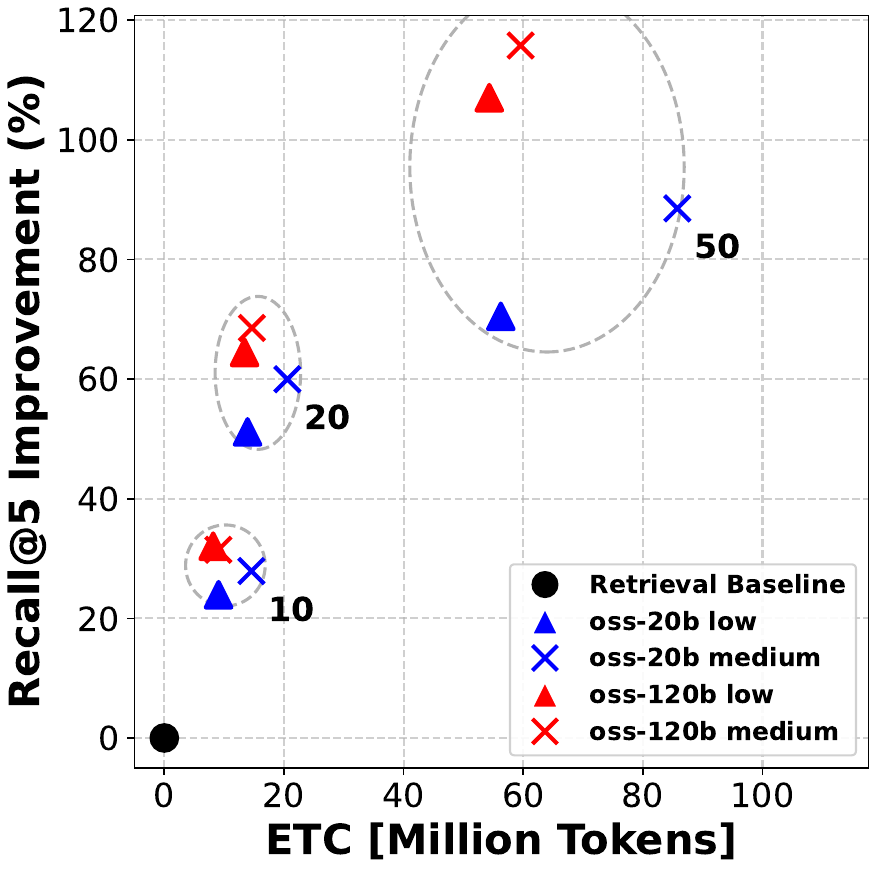}
        \caption{\textbf{$\alpha=0.3, \beta=5$}}
    \end{subfigure}
    \hfill
    \begin{subfigure}[t]{0.32\textwidth}
        \centering
        \includegraphics[width=\textwidth]{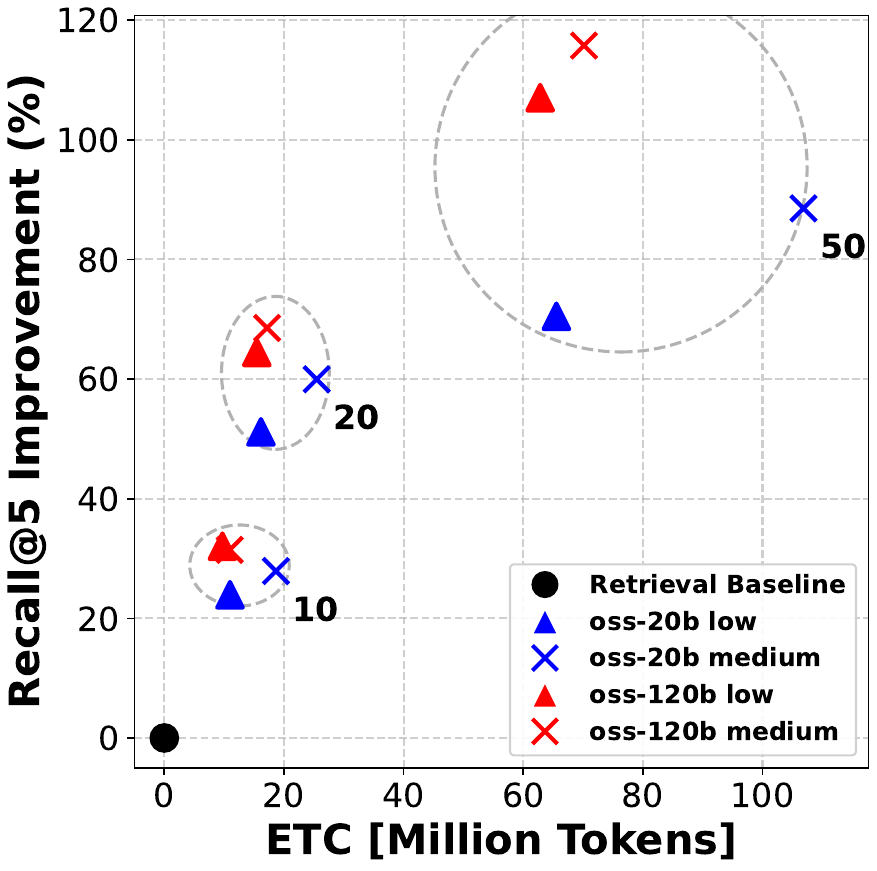}
        \caption{\textbf{$\alpha=0.3, \beta=7$}}
    \end{subfigure}
    \vspace{2mm}
    \begin{subfigure}[t]{0.32\textwidth}
        \centering
        \includegraphics[width=\textwidth]{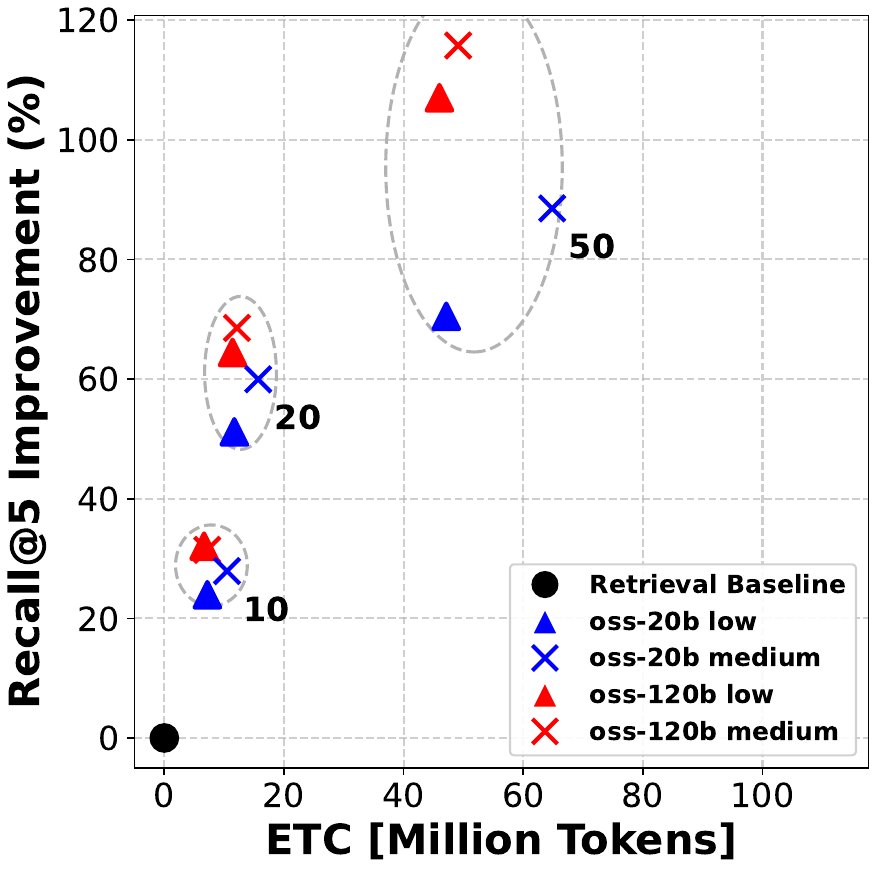}
        \caption{\textbf{$\alpha=0.5, \beta=3$}}
    \end{subfigure}
    \hfill
    \begin{subfigure}[t]{0.32\textwidth}
        \centering
        \includegraphics[width=\textwidth]{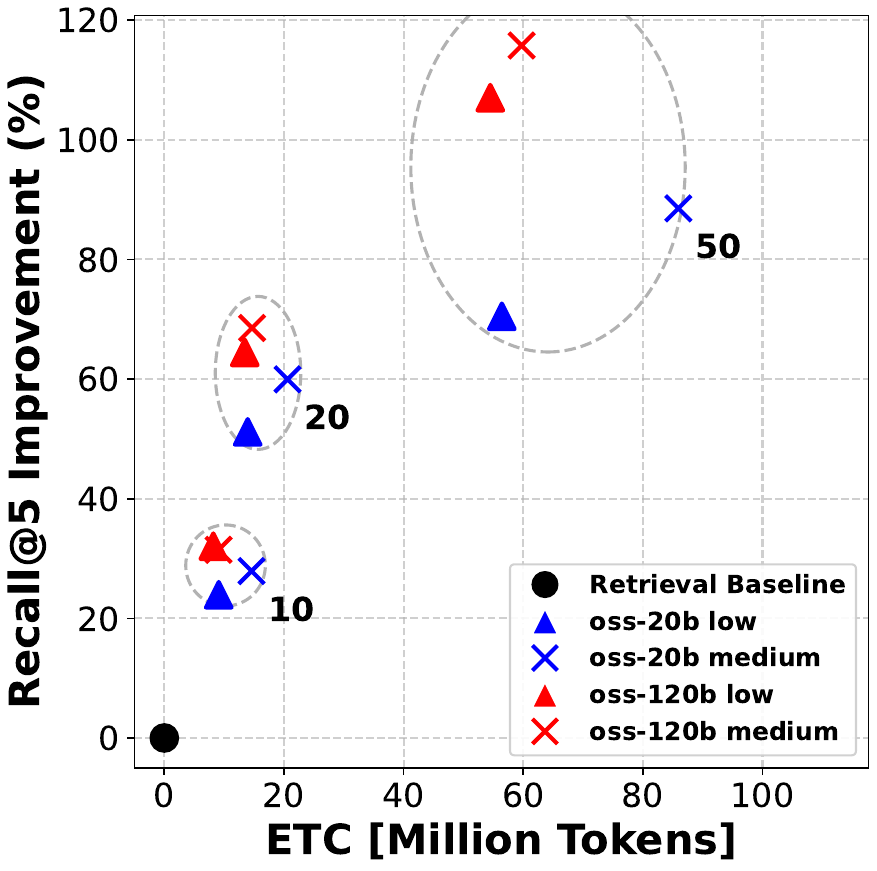}
        \caption{\textbf{$\alpha=0.5, \beta=5$}}
    \end{subfigure}
    \hfill
    \begin{subfigure}[t]{0.32\textwidth}
        \centering
        \includegraphics[width=\textwidth]{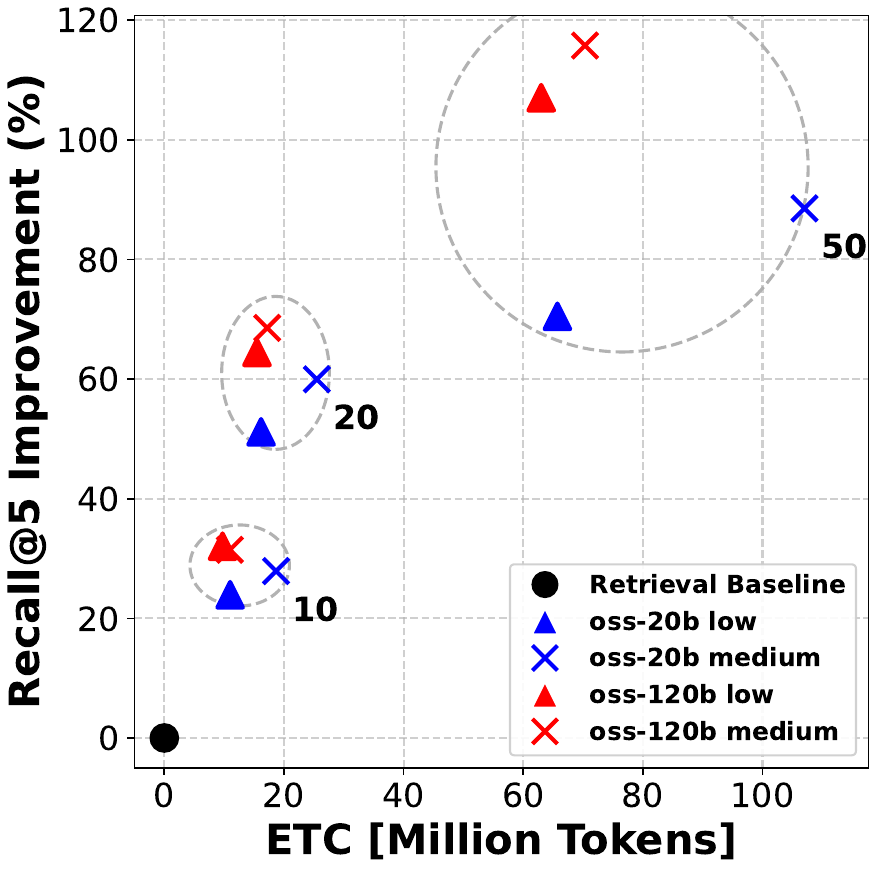}
        \caption{\textbf{$\alpha=0.5, \beta=7$}}
    \end{subfigure}
    \caption{Recall@5 improvement vs. the effective cost per million tokens for reranking depth of $d\in \{10, 20, 50\}$ with oss-20b and oss-120b models under low and medium reasoning efforts.}
    \label{fig:recall_vs_etc_all}
\end{figure*}
\begin{figure*}[t]
    \centering
    \begin{subfigure}[t]{0.32\textwidth}
        \centering
        \includegraphics[width=\textwidth]{figures/etc_Acc_120b_Avg_a0.1_b3_g1.0.pdf}
        \caption{\textbf{$\alpha=0.1, \beta=3$}}
    \end{subfigure}
    \hfill
    \begin{subfigure}[t]{0.32\textwidth}
        \centering
        \includegraphics[width=\textwidth]{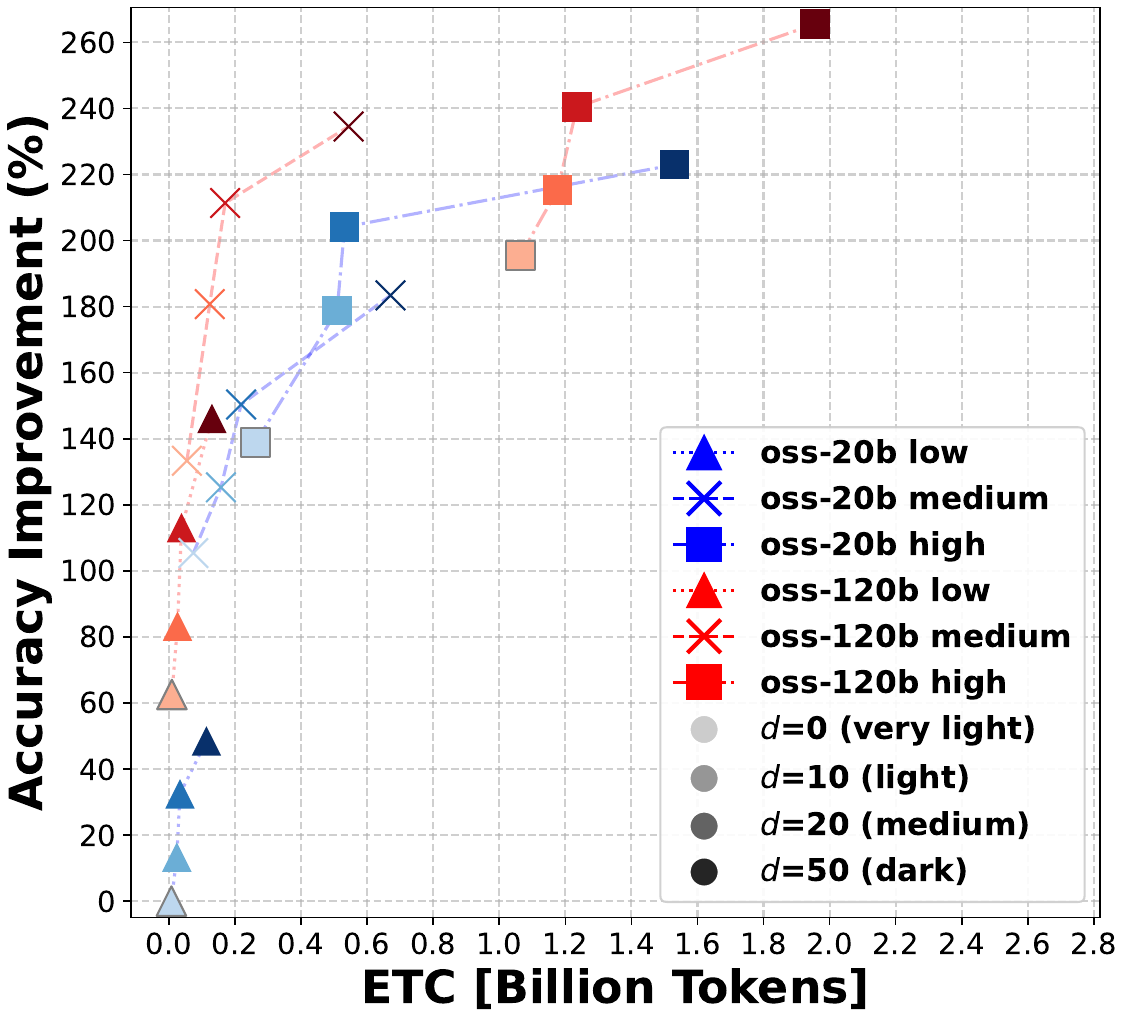}
        \caption{\textbf{$\alpha=0.1, \beta=5$}}
    \end{subfigure}
    \hfill
    \begin{subfigure}[t]{0.32\textwidth}
        \centering
        \includegraphics[width=\textwidth]{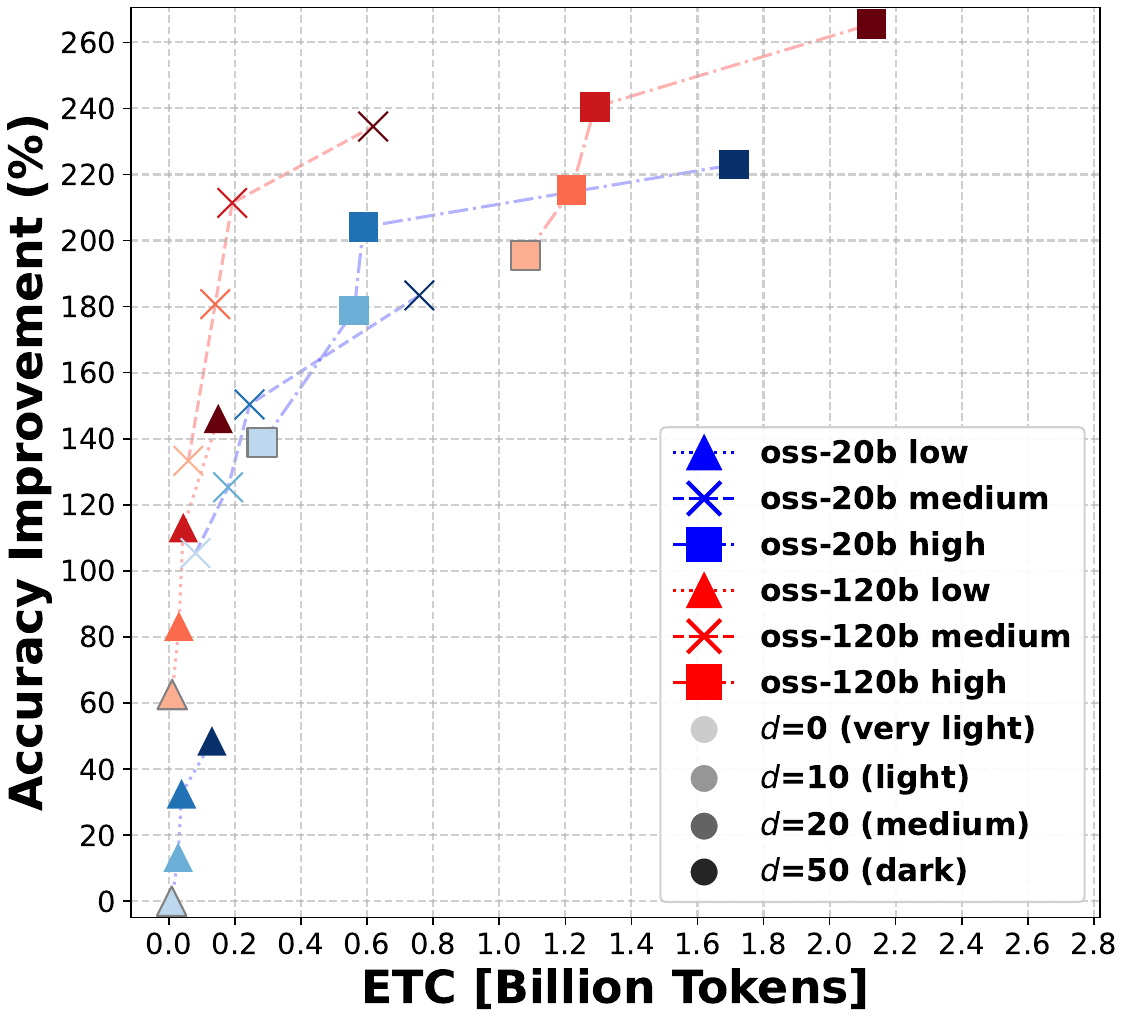}
        \caption{\textbf{$\alpha=0.1, \beta=7$}}
    \end{subfigure}
    \vspace{2mm}
    \begin{subfigure}[t]{0.32\textwidth}
        \centering
        \includegraphics[width=\textwidth]{figures/etc_Acc_120b_Avg_a0.3_b3_g1.0.pdf}
        \caption{\textbf{$\alpha=0.3, \beta=3$}}
    \end{subfigure}
    \hfill
    \begin{subfigure}[t]{0.32\textwidth}
        \centering
        \includegraphics[width=\textwidth]{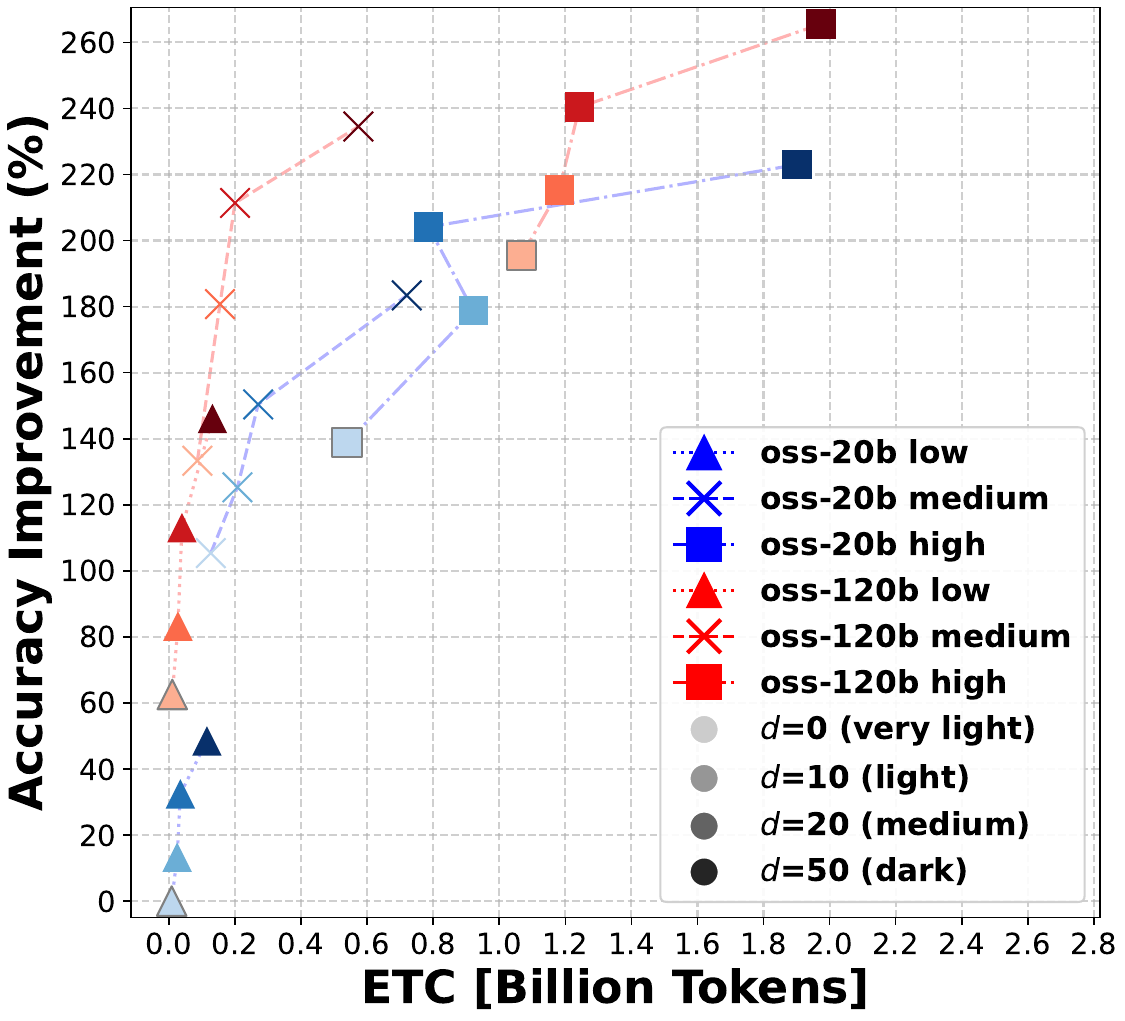}
        \caption{\textbf{$\alpha=0.3, \beta=5$}}
    \end{subfigure}
    \hfill
    \begin{subfigure}[t]{0.32\textwidth}
        \centering
        \includegraphics[width=\textwidth]{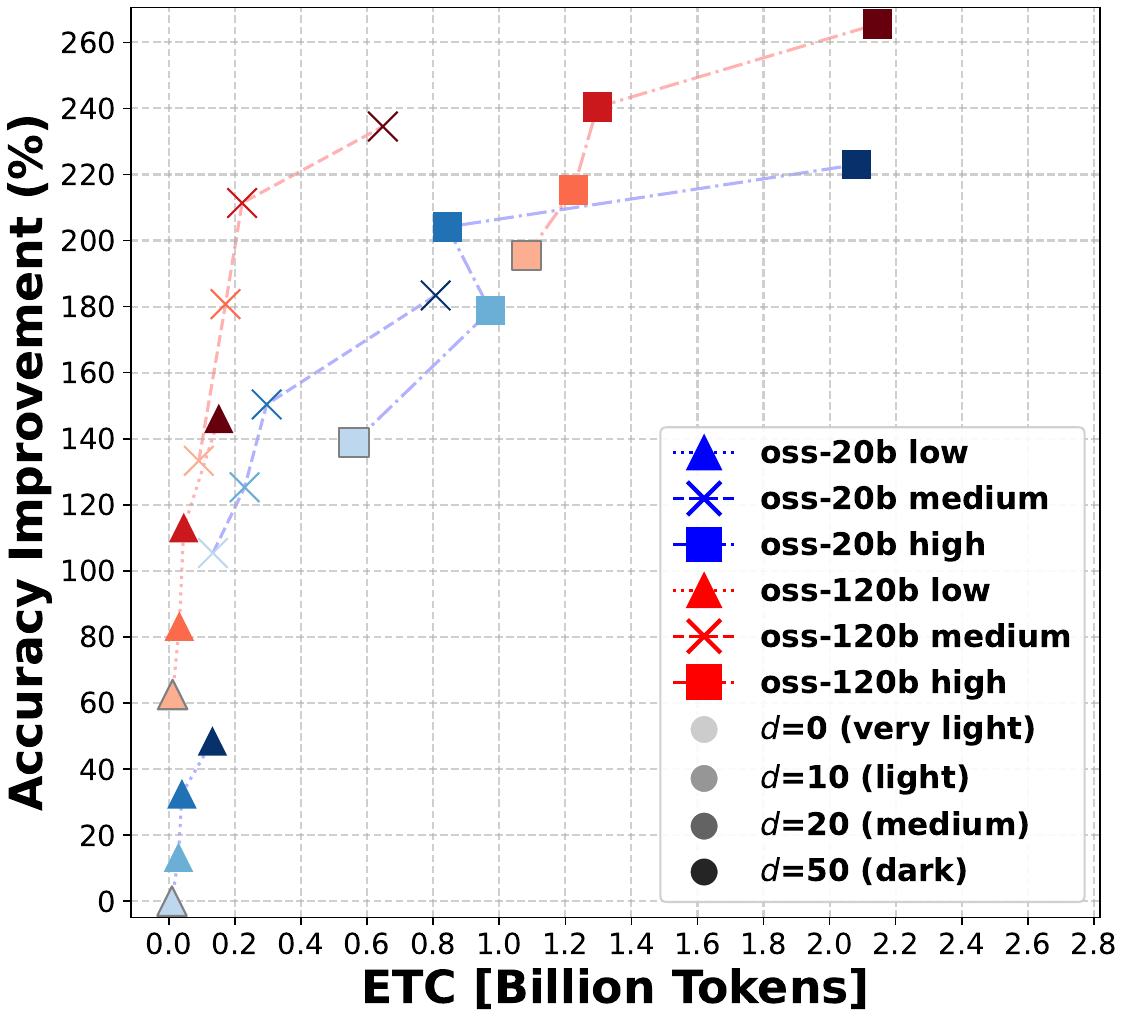}
        \caption{\textbf{$\alpha=0.3, \beta=7$}}
    \end{subfigure}
    \vspace{2mm}
    \begin{subfigure}[t]{0.32\textwidth}
        \centering
        \includegraphics[width=\textwidth]{figures/etc_Acc_120b_Avg_a0.5_b3_g1.0.pdf}
        \caption{\textbf{$\alpha=0.5, \beta=3$}}
    \end{subfigure}
    \hfill
    \begin{subfigure}[t]{0.32\textwidth}
        \centering
        \includegraphics[width=\textwidth]{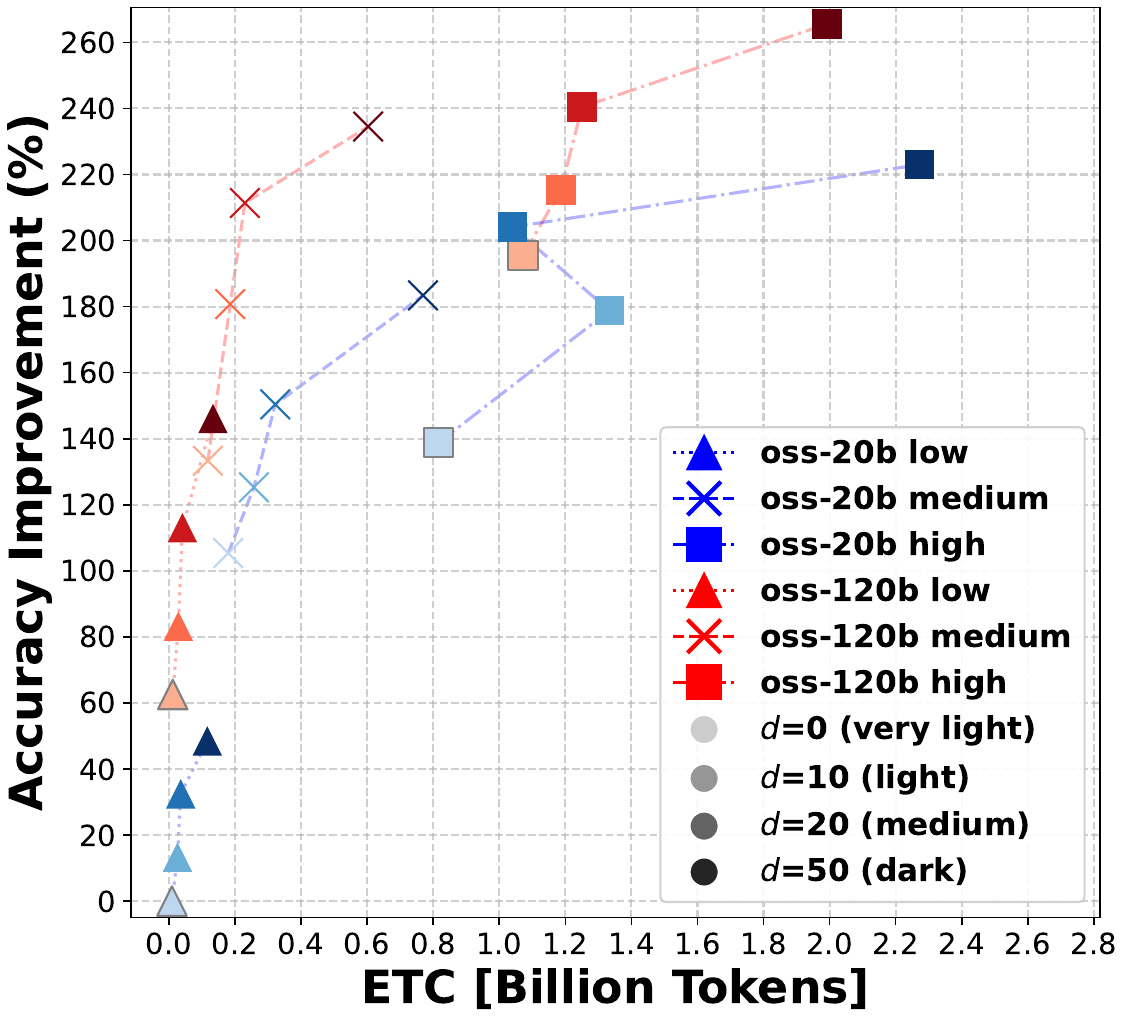}
        \caption{\textbf{$\alpha=0.5, \beta=5$}}
    \end{subfigure}
    \hfill
    \begin{subfigure}[t]{0.32\textwidth}
        \centering
        \includegraphics[width=\textwidth]{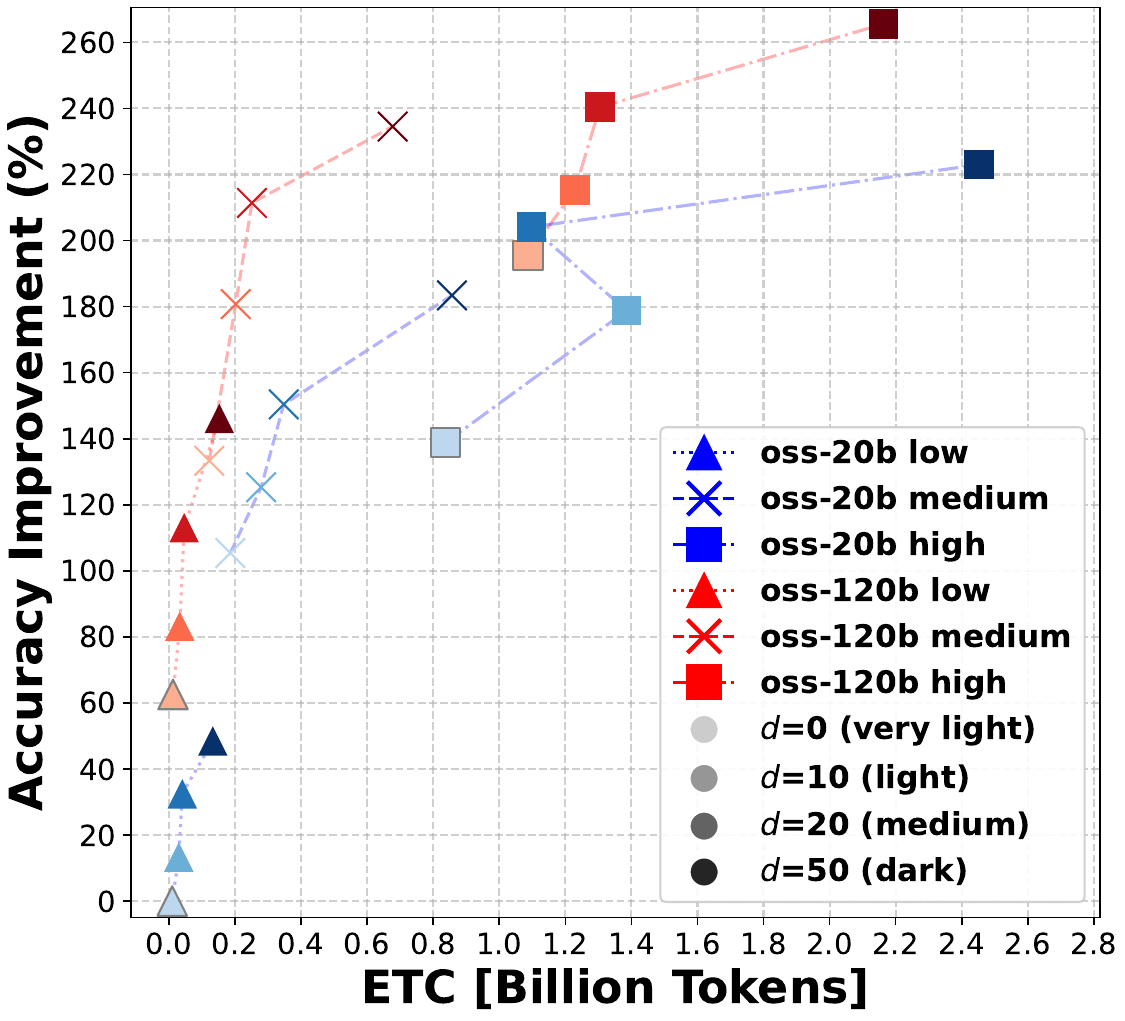}
        \caption{\textbf{$\alpha=0.5, \beta=7$}}
    \end{subfigure}
    \caption{Accuracy improvement vs. the effective cost per ten million tokens for oss-20b and oss-120b deep search agents under low, medium, and high reasoning effort and reranking depth of $d\in \{0, 10, 20, 50\}$. In all cases top-5 candidates are passed to the search agent.}
    \label{fig:accuracy_vs_etc_all}
\end{figure*}
\begin{figure*}[t]
    \centering
    \begin{subfigure}[t]{0.32\textwidth}
        \centering
        \includegraphics[width=\textwidth]{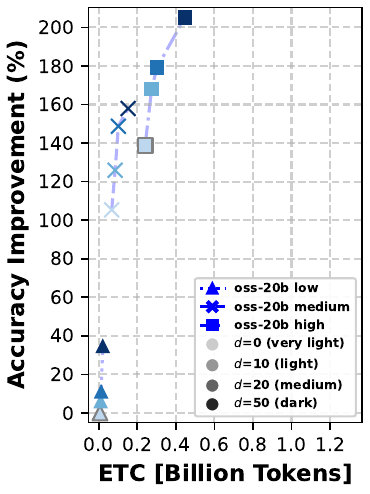}
        \caption{\textbf{$\alpha=0.1, \beta=3$}}
    \end{subfigure}
    \hfill
    \begin{subfigure}[t]{0.32\textwidth}
        \centering
        \includegraphics[width=\textwidth]{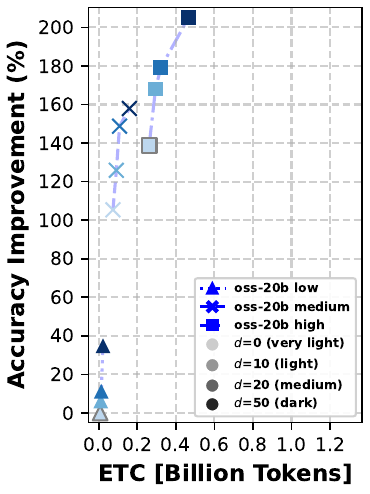}
        \caption{\textbf{$\alpha=0.1, \beta=5$}}
    \end{subfigure}
    \hfill
    \begin{subfigure}[t]{0.32\textwidth}
        \centering
        \includegraphics[width=\textwidth]{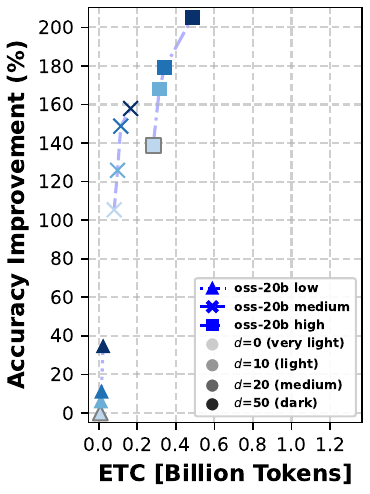}
        \caption{\textbf{$\alpha=0.1, \beta=7$}}
    \end{subfigure}
    \vspace{2mm}
    \begin{subfigure}[t]{0.32\textwidth}
        \centering
        \includegraphics[width=\textwidth]{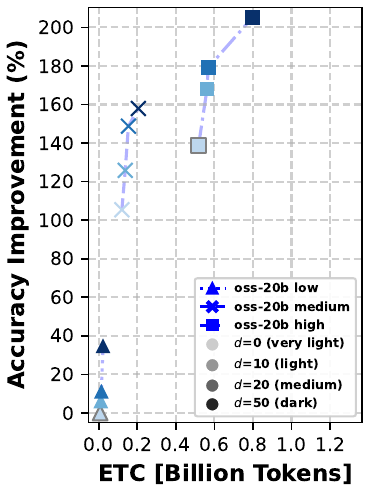}
        \caption{\textbf{$\alpha=0.3, \beta=3$}}
    \end{subfigure}
    \hfill
    \begin{subfigure}[t]{0.32\textwidth}
        \centering
        \includegraphics[width=\textwidth]{figures/etc_Acc_120b_Avg_a0.3_b5_g0.32.pdf}
        \caption{\textbf{$\alpha=0.3, \beta=5$}}
    \end{subfigure}
    \hfill
    \begin{subfigure}[t]{0.32\textwidth}
        \centering
        \includegraphics[width=\textwidth]{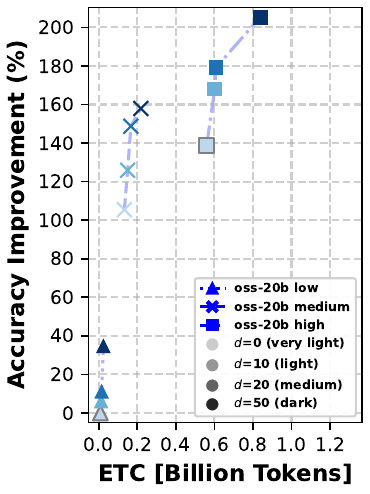}
        \caption{\textbf{$\alpha=0.3, \beta=7$}}
    \end{subfigure}
    \vspace{2mm}
    \begin{subfigure}[t]{0.32\textwidth}
        \centering
        \includegraphics[width=\textwidth]{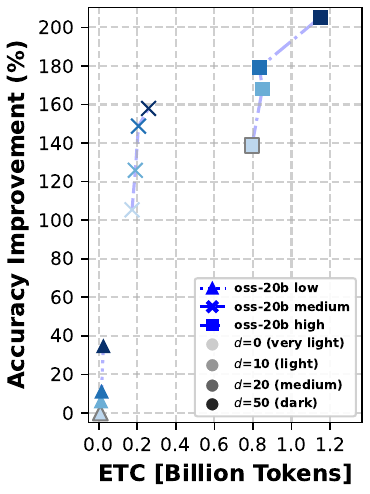}
        \caption{\textbf{$\alpha=0.5, \beta=3$}}
    \end{subfigure}
    \hfill
    \begin{subfigure}[t]{0.32\textwidth}
        \centering
        \includegraphics[width=\textwidth]{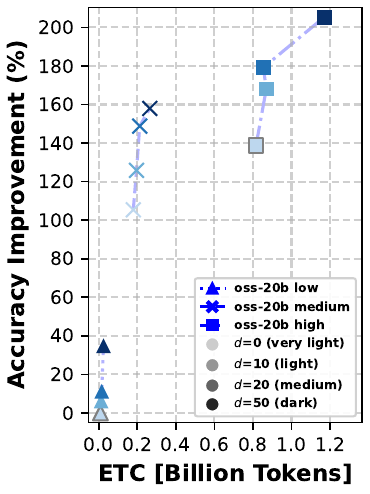}
        \caption{\textbf{$\alpha=0.5, \beta=5$}}
    \end{subfigure}
    \hfill
    \begin{subfigure}[t]{0.32\textwidth}
        \centering
        \includegraphics[width=\textwidth]{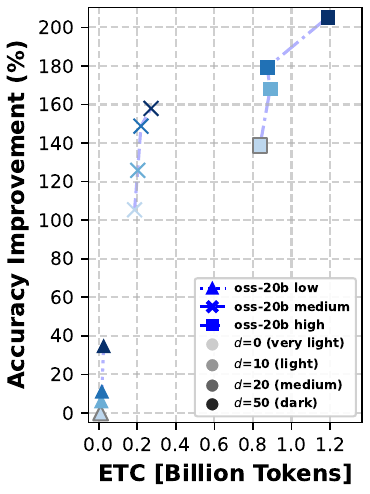}
        \caption{\textbf{$\alpha=0.5, \beta=7$}}
    \end{subfigure}
    \caption{Accuracy improvement vs. the effective cost per ten million tokens for the oss-20b deep search agent under low, medium, and high reasoning effort and reranking depth of $d\in \{0, 10, 20, 50\}$ with qwen3-reranker-0.6b and $\gamma=0.32$. In all cases top-5 candidates are passed to the search agent.}
    \label{fig:accuracy_vs_etc_all_pointwise}
\end{figure*}
\Cref{fig:recall_vs_etc_all} shows the percentage improvement in Recall@5 as a ETC per one million tokens. For a fixed output premium $\beta$, all values of the cached-token discount $\alpha$ yield nearly identical trends, since the vast majority of input tokens are non-cached.

\Cref{fig:accuracy_vs_etc_all} shows the percentage improvement in accuracy as a function of ETC per ten million tokens. For a fixed cached-token discount $\alpha$, all values of the output premium $\beta$ yield nearly identical trends, since both for reranking and search the vast majority of the consumed tokens are input tokens. 
Looking at each column, a notable pattern emerges when cached tokens are discounted less aggressively (i.e., \(\alpha \ge 0.3\)): enabling reranking with moderate depths (\(d = 20\)) can shift the tradeoff curve up and to the left, improving accuracy while maintaining comparable---and in some settings even lower---ETC relative to the previous reranking depth.

\Cref{fig:accuracy_vs_etc_all_pointwise} shows the same analysis for the heterogeneous setup. Due to the lower relative inference FLOPs of the cross-encoder reranker, the gains are more pronounced at a lower ETC increase.

\section{Cross-Run Variance}
\label{app:cross-run variance}
To evaluate the robustness of the \texttt{oss-120b-med} search agent, we conduct five independent trials (see \Cref{tab:run_variance}). Although reranking is deterministic ($T=0$), we observe only modest variance in recall ($58.85\% \pm 0.94$), indicating that stochasticity in the upstream query generation stage has limited impact on the retrieved set. 
While the final reasoning stage operates at the default temperature ($T=1.0$) to encourage diverse query generation, end-to-end accuracy remains stable, with a narrow confidence interval of $52.27\% \pm 1.45$. These results suggest that the agent is highly consistent: the slight increase in variance from recall to accuracy is expected due to the compounding effects of multi-step reasoning, yet the absolute deviation remains sufficiently small to ensure reliable performance across independent runs.

\section{Inference FLOPs/Token Estimation}
\label{app:gflops}

Extending the ETC metric to a heterogeneous setting with different reranker and search agent models requires estimating their relative inference cost in GFLOPs per token. \Cref{tab:arch} summarizes the architectural parameters extracted from publicly available configuration files.\footnote{\url{https://huggingface.co/Qwen/Qwen3-Reranker-0.6B/blob/main/config.json}, \url{https://huggingface.co/openai/gpt-oss-20b/blob/main/config.json}.}

Applying Equation~\ref{eq:flops_total} yields estimated computational costs of 2.58 GFLOPs/token for qwen3-reranker-0.6b and 7.96 GFLOPs/token for gpt-oss-20b, resulting in a normalization factor of $\gamma = 2.58/7.96 \approx 0.32$.

\paragraph{FLOPs Calculation.}
Inference cost is estimated by counting the floating-point multiply-accumulate operations required per decoded token under KV-cache decoding. For a transformer with $L$ layers, the per-layer cost is decomposed into projection, attention, and feed-forward components, defined as follows:
\vspace{-0.3cm}
\begin{align}
\label{eq:flops_total}
\text{FLOPs/token} &\approx \sum_{\ell=1}^{L}
\left(
F^{\text{proj}}_\ell +
F^{\text{attn}}_\ell +
F^{\text{mlp}}_\ell
\right) \\
F^{\text{proj}}_\ell &= 4d^2 \cdot (1 + r_{\text{attn}}) \\
F^{\text{attn}}_\ell &= 2d \cdot T_\ell^{\text{eff}} \\
F^{\text{mlp}}_\ell &= 2k \cdot \alpha \cdot d \cdot d_{\text{ff}}
\end{align}
\noindent where:
\begin{itemize}[ nosep]
    \item $d$: hidden dimension.
    \item $r_{\text{attn}} = n_{\text{kv}} / n_{\text{q}}$: grouped-query attention factor, which equals 1 in standard multi-head attention.
    \item $T_\ell^{\text{eff}}$: effective context length at layer $\ell$, defined as:
\end{itemize}
\begin{equation}
T_\ell^{\text{eff}} =
\begin{cases}
T, & \text{full-attention layers} \\
W, & \text{sliding-window layers}
\end{cases}
\vspace{-0.1cm}
\end{equation}
\begin{itemize}[nosep]
    \item $k$: number of active experts per token ($k=1$ for dense models).
    \item $\alpha$: activation factor ($\alpha=3$ for SwiGLU, $\alpha=2$ for GELU).
    \item $d_{\text{ff}}$: feed-forward or expert dimension.
\end{itemize}
\vspace{0.1cm}

\noindent The projection term accounts for query, key, value, and output linear transformations. The attention term scales linearly with effective context length due to KV-cache decoding, eliminating quadratic dependence. The MLP term accounts only for active parameters, enabling direct applicability to Mixture-of-Experts (MoE) architectures.
\begin{table}[tbh]
\centering
\caption{Reranking token usage ($10^6$) in the one-shot retrieval setting, where the full question is used as the query. Total usage is decomposed into Input, Cache, Output, and Reasoning. Cache denotes input tokens served from cache, and Reasoning denotes output tokens used for reasoning.}
\label{tab:retrieval_token_stats}
\vspace{-0.2cm}
\setlength{\tabcolsep}{5pt}

\resizebox{0.85\columnwidth}{!}{
\begin{tabular}{l c
r r r r r}
\toprule

\textbf{Reranker} & \textbf{d}
& \textbf{Inp.} & \textbf{Cache}
& \textbf{Out.} & \textbf{Reas.}
& \textbf{Total} \\

\midrule

(1a) oss-20b-low   & 10 & 4.40 & 0.13 & 0.95 & 0.92 & 5.35 \\
(1b) oss-20b-med   & 10 & 4.40 & 0.13 & 2.05 & 2.01 & 6.45 \\
(1c) oss-120b-low  & 10 & 4.40 & 0.13 & 0.77 & 0.73 & 5.17 \\
(1d) oss-120b-med  & 10 & 4.40 & 0.13 & 0.95 & 0.91 & 5.35 \\

\midrule

(2a) oss-20b-low   & 20 & 8.41 & 0.13 & 1.12 & 1.07 & 9.53 \\
(2b) oss-20b-med   & 20 & 8.41 & 0.13 & 2.45 & 2.39 & 10.86 \\
(2c) oss-120b-low  & 20 & 8.41 & 0.13 & 1.02 & 0.95 & 9.43 \\
(2d) oss-120b-med  & 20 & 8.41 & 0.13 & 1.27 & 1.19 & 9.68 \\

\midrule

(3a) oss-20b-low   & 50 & 33.62 & 0.85 & 4.65 & 4.44 & 38.26 \\
(3b) oss-20b-med   & 50 & 33.63 & 0.85 & 10.55 & 10.29 & 44.18 \\
(3c) oss-120b-low  & 50 & 33.64 & 0.85 & 4.26 & 3.99 & 37.90 \\
(3d) oss-120b-med  & 50 & 33.64 & 0.85 & 5.30 & 5.01 & 38.94 \\

\bottomrule
\end{tabular}
}
\end{table}

\begin{table}[tbh]
\centering
\caption{Variance analysis of the oss-120b-med search agent with reranking $d=20$ across 5 independent runs. The final top-5 candidates are passed to the agent. Accuracy (Acc.) and Calibration Error (Calib.) report means with 95\% CIs. }
\vspace{-0.2cm}
\label{tab:run_variance}
\resizebox{0.8\columnwidth}{!}{
\begin{tabular}{lrrrrr}
\toprule
\textbf{Run} & \textbf{Srs.} & \textbf{Recall} & \textbf{Acc. (CI)} & \textbf{Calib. (CI)} \\
\midrule
Run 1 & 10.38 & 59.61 & 53.96 (0.07) & 34.77 (0.11) \\
Run 2 & 10.25 & 59.70 & 52.29 (0.00) & 34.93 (0.00) \\
Run 3 & 10.06 & 58.17 & 50.74 (0.16) & 36.90 (0.36) \\
Run 4 & 10.10 & 58.13 & 52.51 (0.07) & 33.74 (0.05) \\
Run 5 & 10.06 & 58.66 & 51.86 (0.17) & 35.42 (0.28) \\
\midrule
\textbf{Average} & \textbf{10.17} & \textbf{58.85 (0.94)} & \textbf{52.27 (1.45)} & \textbf{35.15 (1.44)} \\
\bottomrule
\end{tabular}
}
\end{table}
\begin{table}[tbh]
\centering
\caption{Architectural parameters used for FLOPs estimation of qwen3-0.6b-reranker and gpt-oss-20b models. $*$The max context length of 32k is used for serving the reranker model with vLLM.}
\vspace{-0.2cm}
\label{tab:arch}
\resizebox{\columnwidth}{!}{
\begin{tabular}{lcc}
\toprule
Parameter & qwen3-0.6b-reranker & gpt-oss-20b \\
\midrule
Layers $L$ & 28 & 24 \\
Hidden size $d$ & 1{,}024 & 2{,}880 \\
FFN / expert size $d_{\text{ff}}$ & 3{,}072 & 2{,}880 \\
Attention type & GQA ($n_{\text{q}}=16,\, n_{\text{kv}}=8$) & GQA ($n_{\text{q}}=64,\, n_{\text{kv}}=8$) \\
$r_{\text{attn}}$ & 0.5 & 0.125 \\
MoE experts ($k$/total) & Dense ($k=1$) & $k=4/32$ \\
Activation & SwiGLU ($\alpha=3$) & SwiGLU ($\alpha=3$) \\
Layer pattern & full attention & 12 full + 12 sliding ($W=128$) \\
Max context length $T$ & 32{,}768$^{*}$ & 131{,}072 \\
\bottomrule
\end{tabular}
}
\end{table}
\section{Use of LLMs}
During manuscript preparation, ChatGPT and Gemini were employed to assist with language refinement, grammatical correction, formatting of selected figures and tables, and the generation of short scripts for aggregating experimental results.

\end{document}